\documentstyle[11pt,aaspp4,flushrt,tighten]{article}
\singlespace
\newcommand{\calP}{{\cal P}}
\def\kms{\,{\rm km\,s^{-1}}}
\def\kmsmpc{\,{\rm km\,s^{-1}\,Mpc^{-1}}}

\def\lumunits{\,{\rm ergs\,s^{-1}\,Hz^{-1}}}

\def\msun{\,{\rm M_\odot}}

\def\sfrd{\,{\rm M_\odot\,yr^{-1}\,Mpc^{-3}}}
\def\rd{\,{\rm yr^{-1}\,Mpc^{-3}}}

\def\etal{{et al.\ }}
\def\AB{{\rm AB}}

\def\spose#1{\hbox to 0pt{#1\hss}}
\def\lta{\mathrel{\spose{\lower 3pt\hbox{$\mathchar"218$}}
     \raise 2.0pt\hbox{$\mathchar"13C$}}}
\def\gta{\mathrel{\spose{\lower 3pt\hbox{$\mathchar"218$}}
     \raise 2.0pt\hbox{$\mathchar"13E$}}}

\newcommand{\mincir}{\raise -2.truept\hbox{\rlap{\hbox{$\sim$}}\raise5.truept
\hbox{$<$}\ }}
\newcommand{\magcir}{\raise -2.truept\hbox{\rla669p{\hbox{$\sim$}}\raise5.truept
\hbox{$>$}\ }}
\newcommand{\minmag}{\raise-2.truept\hbox{\rlap{\hbox{$<$}}\raise 6.truept\hbox
{$>$}\ }}

\newcommand{\be}{\begin{equation}}
\newcommand{\ee}{\end{equation}}
\newcommand{\ba}{\begin{eqnarray}}
\newcommand{\ea}{\end{eqnarray}}
\newcommand{\brr}{\begin{array}}
\newcommand{\err}{\end{array}}
\newcommand{\bc}{\begin{center}}
\newcommand{\ec}{\end{center}}
\newcommand{\f}{\frac}
\newcommand{\Om}{\Omega}
\newcommand{\de}{\delta}
\newcommand{\s}{\sigma}

\begin{document}
\title{Gravitational Lensing of Distant Supernovae in Cold Dark Matter 
Universes}
\author{Cristiano Porciani\altaffilmark{1,2} and Piero Madau\altaffilmark{1,3}}

\altaffiltext{1}{Space Telescope Science Institute, 3700 San Martin 
Drive, Baltimore, MD 21218.}
\altaffiltext{2}{Scuola Internazionale Superiore di Studi Avanzati, 
via Beirut 2-4, 34014 Trieste, Italy.} 
\altaffiltext{3}{Institute of Astronomy, Madingley Road, 
Cambridge CB3 0HA, UK.}

\begin{abstract}
\noindent Ongoing searches for supernovae (SNe) at cosmological 
distances have recently started to provide large numbers of events with 
measured redshifts and apparent brightnesses. Compared to quasars or 
galaxies, Type~Ia SNe represent a population of sources with well-known 
intrinsic properties, and could be 
used to detect gravitational lensing even in the absence of multiple or highly 
distorted images. We investigate the lensing effect of background SNe due to 
mass condensations in three popular cold dark matter cosmologies ($\Lambda$CDM,
OCDM, SCDM), and compute lensing frequencies, rates of SN explosions, and 
distributions of arrival time differences and image separations. If dark halos 
approximate singular isothermal spheres on galaxy scales 
and Navarro-Frenk-White profiles on group/cluster scales, and are distributed 
in mass 
according to the Press-Schechter theory, then about one every 12 SNe at 
$z\sim 1$ will be magnified by $\Delta m\ge 0.1$ mag (SCDM). 
The detection rate of 
SN~Ia with magnification $\Delta m\ge 0.3$ is estimated to be of order a few 
events yr$^{-1}$ deg$^{-2}$ at maximum $B$-light and $I_\AB\le 25$, a hundred 
time smaller than the total rate expected at these magnitude levels. In the 
field, events magnified by more than 0.75 mag are 7 times less frequent: 
about one fifth of them gives rise to observable multiple images. While the 
time delay between the images is shorter than 3 days (30 days) in $\sim 25\%$ 
(50\%) of the cases (SCDM), a serious bias against the detection of 
small-separation events in ground-based surveys is caused by the luminosity 
of the foreground lensing galaxy. 
Because of the flat $K$-correction and wide luminosity function, Type II 
SNe dominate the number counts at $I_\AB>25$ and have the largest fraction of
lensed objects. The optimal survey sensitivity for Type Ia's magnified by 
$\Delta m\ge 0.75$ mag is $I_\AB\approx 23$. Magnification bias increases 
their incidence by a factor of 50 in samples with 
$I_\AB\le 22$, dropping to a factor of 3.5 at 24 mag. At faint magnitudes the 
enhancement is larger for SN II.

\end{abstract}

% \twocolumn

\section{Introduction}

The potential of gravitational lensing as a tool for the determination of 
cosmological parameters or as a probe of cosmogonic models has long been
recognized (see Blandford \& Narayan 1992 for a review). Frequently discussed 
techniques include detailed models of specific lens systems such as multiple 
quasars (e.g. Falco, Gorenstein, \& Shapiro 1991; Grogin \& Narayan 1996; 
Kundic \etal 1997), arcs (Lynds \& Petrosian 1989; Tyson \& Fisher 1995; 
Tyson, Kochanski, \& Dell' 
Antonio 1998), and radio rings (Kochanek \etal 1989), 
statistical studies of lensed QSOs (Narayan \& White 1988; Fukugita \etal 1992; 
Kochanek 1996; Wambsganss \etal 1995), and measurements of the correlated 
ellipticities induced by large-scale structure in the images of background
galaxies (Babul \& Lee 1991; Blandford \etal 1991; Kaiser 1992; Villumsen 
1996).  

It has been recently realized by many authors (Kolatt \& Bartelmann 1998; 
Metcalf 1999; Holz 1998; Marri \& Ferrara 1998; Wang 1998; see also Linder,
Schneider, \& Wagoner 1988) that gravitational
lensing of distant supernovae (SNe) may provide a new tool for doing 
cosmology. When corrected for light and color curve shapes, Type Ia SNe can 
be calibrated to be excellent standard candles, with a brightness at 
maximum $B$-light of $M_V=-19.4\pm 0.15$ mag (Riess, Press, \& Kirshner
1996; Hamuy \etal 1996). Systematic searches  for SNe at high redshifts
(Perlmutter \etal 1998; Garnavich \etal 1998) have already yielded a plethora
of events. At this rate of accumulation, several hundreds of high-$z$ SNe will 
be known within the next few years. The uniformity of Type Ia's makes them 
unique sources for measuring the redshift-luminosity distance relation.

In the observable (clumpy) universe, gravitational lensing will magnify and 
demagnify high redshift supernovae relative to the predictions of ideal 
(homogeneous) reference cosmological models. The magnitude and frequency of the 
effect depend on the 
power spectrum of density fluctuations, the abundance and the clustering 
properties of virialized clumps, the mass distribution within individual lenses,
and the underlying world model. The  degree to which weak lensing increases the 
level of noise in the Hubble diagram of Type Ia SNe, thereby decreasing the 
precision of cosmological parameter determinations, has been discussed by 
Kantowski, Vaughan, \& Branch (1995), Frieman (1996), Wambsganss \etal (1997), 
Holz (1998), and Metcalf (1999). The present paper examines instead the 
high magnification tail of the lensing distribution. The likelihood of a 
SN at $z_s\lta 2$ being magnified by $\Delta m\ge 0.3$ ($\ge 0.1$) mag by a galaxy, 
group, or cluster is small but non-negligible, $\sim {\rm few}\, \times 
10^{-2}$ (10\%). One can then use recent estimates of the global history of 
star formation to compute the expected frequencies of lensed SNe as a function 
of cosmic time. The detection rate of Type Ia's with magnification $\ge 0.3$ 
mag is estimated to be of order a few SNe yr$^{-1}$ deg$^{-2}$ at 
$I_\AB\le 25$ mag. Events with $\Delta m\ge 0.75$ mag
are seven times less frequent (unless 
specifically targeted, Kolatt \& Bartelmann 1998). At bright magnitudes, the 
effect of magnification bias on the apparent frequency of lensed SNe is huge. 

On the face of their rarity, which makes a measurement of the 
magnification probability density (hence of the distribution of 
mass inhomogeneities in the universe) a 
very difficult task, the modeling of individual lensed SN~Ia might have
a few advantages over the study of multiple quasars or galaxy correlated 
ellipticities. Lensed SNe will appear unusually bright for their redshift 
and will be easily identifiable by the proximity of a foreground galaxy, 
group, or cluster. Their magnification, which is related to the integrated 
mass density along the line-of-sight, is 
directly measurable to a good precision. A supernova that goes off within the 
Einstein ring of a foreground mass concentration may generate multiple
images at different positions on the sky. The time delay between the images, 
inversely proportional to the Hubble constant, is longer than a month in about 
$45-55\%$ of the cases, and could be measured by comparing the light curves of 
the two events during a reasonable monitoring period. If all the relevant 
deflection angles to constrain the lens system were known, it could be 
possible to estimate $H_0$. Small-separation events are associated with 
galaxy lenses, and will remain undetected because of the presence of such a 
luminous foreground object. 

The plan of the paper is as follows. In \S~2 we compute the optical depth
for a supernova to undergo a lensing event caused by an intervening dark
matter halo. We investigate three popular hierarchical cosmogonies, two 
flat models with $\Omega_M=0.35$ and 1, and an open model with 
$\Omega_M=0.3$. The population of gravitational lenses is modeled as a 
collection of singular isothermal spheres on galaxy scales and 
Navarro-Frenk-White halos (Navarro, Frenk, \& White 1997; hereafter NFW)
on group/cluster scales, distributed in mass according 
to the Press-Schechter theory (Press \& Schechter 1974; hereafter PS). 
The rates of SN explosions as a function of cosmic time are estimated 
in \S~3, together with the expected number counts of magnified events, 
distributions of time delays and image separations, and selection effects. As 
the rest-frame flux
of Type Ia SNe falls rather steeply below 3500\,\AA\ and we are mostly 
interested in redshifts $z\gta 1$, the counts are computed in the $I$-band. We 
briefly summarize our results in \S~4.  

\section{Basic theory}

If the geometry of the universe is well approximated on large scales by the 
Friedmann-Robertson-Walker (FRW) metric, 
the optical depth for a light-beam emitted by a point source
at redshift $z_s$ and received at $z=0$ through a lensing event is
(Turner, Ostriker, \& Gott 1984; Fukugita \etal 1992)
\be
\tau(z_s)=\int_0^{z_s} dz (1+z)^3 
\f{dl}{dz}
\int_0^\infty dM \,\s(M,z,z_s) n_L(M,z) \;,
\label{tautog}
\ee
where $dl/dz=c H_0^{-1} (1+z)^{-1} [\Om_M(1+z)^3+\Om_K(1+z)^2+\Om_{\Lambda}]^
{-1/2}$ is the cosmological line element, 
$\Om_K=1-\Om_M-\Om_{\Lambda}$ is the curvature contribution to
the present density parameter, $\s(M,z,z_s)$ is the cross section for 
lensing as measured on the lens-plane, and $n_L(M,z)$ denotes the 
comoving differential distribution of lenses at redshift $z$ with respect to 
the variable $M$ (i.e. mass or velocity dispersion). Equation (\ref{tautog}) 
assumes that each bundle of light rays encounters only one lens, the 
lens population to be randomly distributed, and the resulting $\tau\ll 1$. 

The mass distribution in a single lens and the geometry of the
source-lens-observer system completely determine the lensing cross section.
The geometry of the lens system generally affects $\s(M,z,z_s)$ through the 
ratio $D(0,z_l) D(z_l,z_s)/ D(0,z_s)$, with $D(z_1,z_2)$ the angular diameter 
distance between $z_1$ and $z_2$. In a smooth (homogeneous) FRW universe, one 
has:
\be
D(z_1,z_2)=\f{c}{H_0} \f{S(\chi_2-\chi_1)}{\sqrt{\kappa}(1+z_2)}
\label{addflrw} \;,
\ee
where $S(x)={\rm sinh}(x)$ if $\Om_K > 0$,
$S(x)=x$ if $\Om_K =0$, $S(x)=\sin (x)$ if $\Om_K<0$, and
\be
\chi_2-\chi_1=\sqrt{\kappa} \int_{z_1}^{z_2} \f{dz}
{[\Om_M(1+z)^3+\Om_K(1+z)^2+\Om_{\Lambda}]^{1/2}}
\ee
with $\kappa=1$ if $\Om_K=0$ and $\kappa=|\Om_K| $ in all other cases.
Here, the density of matter within a cone of light rays is always the same as 
the mean density in the universe (filled beam approach). The observed 
universe, however, is clumpy on galaxy scales, and local mass concentrations 
are known to produce significant changes in the redshift-distance relation 
(Zel'dovich 1964). Numerical studies based on ray-shooting 
techniques (e.g. Tomita 1998) 
%(Kasai, Futamase \& Takahara 1990; Watanabe \& Tomita 1990; 
%Fukushige \& Makino 1994; Tomita 1998) 
have shown that a definite 
distance-redshift relation does not exist in a clumpy universe.
Analytical approximations are available for light beams that subtend a
small solid angle, when the probability of intercepting a mass concentration
is small and light passes through very underdense regions (empty beam 
approach, Dyer \& Roeder 1973; Ehlers \& Schneider 1986).
For large apertures
(where the mass density in a beam approaches the average one),
it is widely believed that
the angular diameter distance reduces, practically, to the FRW case. 
Note, however, that
caustic formation might alter the FRW distance relation even on large scales
(Ellis, Bassett, \& Dunsby 1999, and references therein).
For simplicity, in the following we will only consider the most widely 
used filled beam approximation. Depending on the underlying cosmology, the 
discrepancies between the lensing frequences obtained within the various 
schemes for light propagation become significant only for $z\gta 2$.

\subsection{Singular isothermal lens}

The standard minimal model to describe the mass distribution
of an extended lens (galaxy or cluster) is the singular isothermal sphere 
(SIS), {\bf $\rho(r)=\s_v^2/2 \pi G r^2$,}  
%
%\be
%\rho(r)=\f{\s_v^2}{2 \pi G} \f{1}{r^2} \;,
%\label{profile}
%\ee
%
where the mass density $\rho$ as a function of the radial coordinate $r$ is 
parametrized through $\s_v$, the one-dimensional velocity dispersion.
%  By projecting 
%the SIS mass distribution along the line-of-sight, one obtains
%
%\be
%\Sigma(\xi)=\f{\s_v^2}{2G} \f{1}{\xi} \;,
%\ee
%
%where $\xi$ is the distance from the center of the two-dimensional profile.
In the 
thin lens approximation (e.g. Schneider, Ehlers \& Falco 1992; Narayan \& 
Bartelmann 1997), this corresponds to a constant deflection angle for incident 
light
rays, $\beta=4 \pi (\s_v/c)^2 =1.4'' \;(\s_v/220\; {\rm km}\;{\rm s}^{-1})^2$,
always pointing towards the lens center of symmetry.  For a source that, in the
absence of lensing, would be seen at an angular distance $\theta$ from this
center, the lens equation leads to a solution (image) at $\theta_+=\theta+
\theta_E$ with magnification $\mu_+=\theta_E/\theta+1$, where $\theta_E=\beta 
D_{ls}
/D_s$ is the Einstein radius [hereafter $D_{ls}\equiv D(z_l,z_s)$, $D_s\equiv 
D(z_s)$ 
and $ D_l\equiv D(z_l)$]. If the alignment is close enough for multiple 
imaging, 
$\theta\le \theta_E$ (strong lensing), a second image is produced at $\theta_-=
\theta-\theta_E$
with magnification $\mu_-=\theta_E/\theta-1$. The separation between the two
images is $2\theta_E$, and their time delay is $c\Delta t=2\theta \theta_E 
(1+z_l) D_l D_s/ D_{ls}$. At a given angle $\theta$, the total magnification 
of the two images is proportional to their separation. A third image, with 
zero magnification, is located in correspondence of the lens center, and 
acquires a finite flux if the singularity that characterizes the lens 
profile is replaced by a core region with finite density. 

A cross section for strong lensing events, $\s_{\rm SIS}$, can be easily 
associated to each SIS lens. Measuring $\s_{\rm SIS}$ in angular units,
we obtain
\be
\s_{\rm SIS}(\s_v,z_s,z_l) = \pi \theta_E^2 =16 \pi^3 \left( 
\f{\s_v}{c}\right)^4 \left( \f{D_{ls}}{D_{s}} \right)^2 \;.
\label{crsecsis}
\ee
The cross section for the image outside the Einstein ring (including both 
multiply and singly imaged objects) being magnified by a 
factor $\mu>1$ can be easily related to $\s_{\rm SIS}$ through
\be
\s(\mu_+>\mu)=\f{\s_{\rm SIS}}{(\mu-1)^2} \;.
\label{sigma+}
\ee
Hence the cross section associated with magnification (say) $\mu=1.316$ 
(corresponding to a brightening of $\sim 0.3$ magnitudes) is $\s(\mu_+>1.316)=10\,
\s_{\rm SIS}$. Weakly magnified events are much more common, $\s(\mu_+>1.1)=
100\, \s_{\rm SIS}$. In the strong lensing regime, the corresponding value 
for the second (dimmest) image ($\mu>0$) is
\be
\s(\mu_->\mu)=\f{\s_{\rm SIS}}{(\mu+1)^2}\;.
\label{sigma-}
\ee
The cross section for a total magnification $\mu_{\rm tot}=\mu_+ + 
\mu_-=2\theta_E/
\theta >\mu$ (with $\mu\ge 2$) is
\be
\s(\mu_{\rm tot}>\mu)= \f{4\s_{\rm SIS}}{\mu^2}\;. 
\ee
Therefore, given a population of sources at $z_s$, the distribution of the 
corresponding magnifications ($\mu_+$, $\mu_-$ and $\mu_{\rm tot}$) observed 
at $z=0$ is directly proportional to the optical depth $\tau(z_s)$ when the 
latter is computed by replacing $\sigma$ with $D_l^2 \sigma_{\rm SIS}$
in equation (\ref {tautog}).

\subsection{Press-Schechter mass function}

We will follow the approach of Narayan \& White (1988) (see also Kochanek
1995b), and use the PS theory for our analysis of the abundance 
of gravitational 
lenses in a cold dark matter (CDM) hierarchical universe. The mass 
distribution of collapsed halos, assembled through accretion and merging
processes, follows then directly from the statistical properties of the 
linear cosmological density fluctuation field, assumed to be Gaussian.  
According to PS, the differential comoving number density of dark halos with 
mass 
$M$ at redshift $z$ is 
\be
n(M,z)=\f{1}{\sqrt{2 \pi}}\f{\rho_0}{M}\f{\de_c(z)}{\s_M^3}\exp\left[-
\f{\de_c^2(z)}{2 \s_M^2} \right] \left| \f{d\s_M^2}{dM}\right |\;,
\label{ps}
\ee
where $\rho_0$ is the present mean density of the universe. The halo abundance 
is 
then fully determined by the linearly extrapolated (to $z=0$) 
variance of the mass-density field smoothed on the scale $M$, $\s_M^2$, and by 
the 
redshift-dependent critical overdensity $\de_c$. The former is determined by
assuming a post-inflationary spectrum of density fluctuations and by following 
its 
linear evolution. 
In a isotropic universe, we can define the power spectrum 
through the two point correlation function in Fourier space
$\langle \delta({\bf k}) \delta({\bf k'})
\rangle = (2 \pi)^3 \delta_D({\bf k}+{\bf k'}) P(k)$,
where $\delta_D$ is the Dirac function.
The variance is then
given by the convolution
\be
\s_M^2=\f{1}{2 \pi^2}\int_0^\infty dk\, k^2 P(k) W^2(kr_0)\;,
\label{var}
\ee
where $W(x)$ is the top-hat filter function. The critical overdensity is 
obtained instead by solving spherical 
top-hat collapse and by computing the linear overdensity corresponding to the 
collapse time. Details of these calculations can be found in Lacey \& Cole 
(1993)
for vanishing cosmological constant, and Eke, Cole \& Frenk (1996) for a general 
flat 
cosmology. We adopt their analytical approach; alternatively one can use 
$N$--body 
simulations of gravitational clustering to 
estimate $\de_c(z)$ as a best fitting parameter (e.g. Lacey \& Cole 1994).

Assuming that every halo virializes to form a (truncated) singular isothermal 
sphere of velocity dispersion $\s_v$, mass conservation implies
\be
\s_v(M,z)=\f{1}{2} H_0 r_0\, \Omega_M^{1/3} \Delta^{1/6} 
\left[ \f{\Omega_M}{\Omega(z)} \right]^{1/6} (1+z)^{1/2} \;,
\label{virial}
\ee
with $\Omega(z)=\Omega_M (1+z)^3/[\Omega_M (1+z)^3+\Omega_K (1+z)^2 
+\Omega_\Lambda]$.
Here $r_0=(3 M/4 \pi \rho_0)^{1/3}$ is the comoving 
Lagrangian (i.e. initial, formally corresponding to infinite redshift) 
radius of the 
collapsing perturbation, $z$ denotes the virialization epoch of the halo,
and $\Delta(z)$ is the ratio between its actual mean density at 
virialization and the corresponding critical density for closure of the
universe. This is determined 
by assuming that the virialization time is twice the turn-around time (in the 
limit $\Omega_M=1$, one finds the standard result $\Delta=18 \pi^2$).  While 
in a universe with $\Om_\Lambda=0$ the radius of a 
virialized spherical perturbation, $r_{\rm vir}$, 
is half its turnaround radius, 
$r_{\rm to}$, in the presence of a cosmological constant the radius at virial 
equilibrium depends on the vacuum contribution to the background energy density, 
and is always smaller that $r_{\rm to}/2$ (Lahav \etal 1991). A detailed fitting 
formula for the ratio $r_{\rm vir}/r_{\rm to}$ as a function of $\Om_\Lambda$ 
and 
$z$ is given in Kochanek (1995b).

Equation (\ref{virial}) relates the PS mass function to the SIS lens profile we 
discussed in the previous section, therefore allowing the computation 
of the optical depth given in equation (\ref {tautog}).

\subsection{Navarro-Frenk-White lens}

The number of known gravitationally lensed quasars has grown to around 50,  
\footnote{see http://cfa-www.harvard.edu/castles.}\, and no confirmed lensed
system has been
detected with angular separation between the images larger than $6.1 ''$.
%\footnote{The possibility that some large separation doubly imaged quasar
%(e.g. Q2345+007 with $\Delta\theta=7''.3$)
%is actually produced by dark lenses
%has, however, been proposed in the literature (e.g. Hawkins 1997).}
It is well known that models for statistical lensing which
combine the PS distribution with SIS profiles
tend to overpredict the fraction of large separation lenses 
($\Delta\theta \gta 6''$)  with respect
to the data, even when selection effects are accounted for 
(e.g. Kochanek 1994; Flores \& Primack 1996).
While photometric 
analyses of E/S0 galaxies performed with the {\it Hubble Space Telescope} ({\it
HST}) (Tremaine \etal 1994) and a number of gravitational lensing studies 
(e.g. Wallington \& Narayan 1993; Kochanek 1995a) indicate that 
a singular isothermal profile is a good approximation for lensing galaxies,
%The giant
%arcs associated with the lensing of background galaxies by galaxy clusters 
%also seem to require cluster ``core'' radii (where the density is 
%approximately constant) to be rather small (Grossman \& Narayan 1988; Soucail 
%\& Mellier 1994).
high-resolution $N$--body simulations have recently shown that virialized
dark matter halos, formed through hierarchical
clustering, have a universal (spherically averaged) 
density profile which is shallower 
than isothermal (but still diverging like $\rho\propto r^{-1}$) 
near the halo center,
and steeper than isothermal (with $\rho\propto r^{-3}$) in its outer regions
(NFW), 
\be
\rho(r)=\f{\rho_{\rm crit} \,\delta_{\rm NFW}}
{(r/r_{\rm s})(1+r/r_{\rm s})^2} \;. 
\ee
Here $r_{\rm s}$ is a scale radius, 
$\delta_{\rm NFW}$ is a dimensionless parameter
which indicates the characteristic density contrast of the halo,
and $\rho_{\rm crit}$ is the critical density of the universe.\footnote{Note, 
however, that 
the slope of the universal density profile in the innermost regions of halos
is still subject of debate. Moore \etal (1999) find  
$\rho\propto r^{-1.5}$, while Kravtsov \etal (1998) argue for the presence
of a central core.}\, Measuring the halo concentration through the 
parameter $C=r_{\rm vir}/r_{\rm s}$, it follows from the definition of 
$\Delta$ that $\delta_{\rm NFW}=(\Delta/3) C^3/[\ln(1+C)-C/(1+C)]$.
The equilibrium density profile of a halo with a given virial radius
is then completely
specified by a single parameter ($\delta_{\rm NFW}$  or $C$).
%In the following discussion, we will adopt the
%simple algorithm described in the appendix of NFW
%to compute the characteristic density contrast
%of hierarchically formed halos.

The lens equation for the NFW density distribution has been computed
by Bartelmann (1996) and Maoz \etal (1997), and can be easily solved
adopting standard numerical techniques.
The lensing efficency of a NFW halo can be parameterized by the
dimensionless quantity $k_s= \delta_{\rm NFW} \rho_{\rm crit} r_{\rm s}/$
$\Sigma_{\rm cr}$, where $\Sigma_{\rm cr}=(c^2/4 \pi G) (D_s/D_{ls} D_l)$
is the critical surface mass density for multiple images.
For $z\sim 0.5$, $z_s \sim 2$ and lens masses between $10^{12}$ and $10^{15}
M_\odot$,   
typical values for $k_s$ lie in the interval $0.07-0.4$.
%In the strong lensing regime a NFW lens (like every axially 
%symmetric lens) produces three images,
%even though one of them always comes out
%strongly demagnified and positioned near the center of the lens.
With respect to SIS profiles containing the same total mass and located
at the same redshift, NFW lenses have much smaller cross sections
for strong lensing (typically $\s_{\rm NFW}$ is a few orders of magnitudes 
smaller than $\s_{\rm SIS}$, the two cross sections differing less
at increasing masses). 
On the other hand, the average magnification of the images is higher for
NFW halos.
Thus, depending on the halo mass (halos 
of increasing mass being less centrally concentrated, NFW), a
SIS approximation
will overestimate the strong lensing optical depth and underestimate 
the total magnification. 

\begin{figure}
\plotone{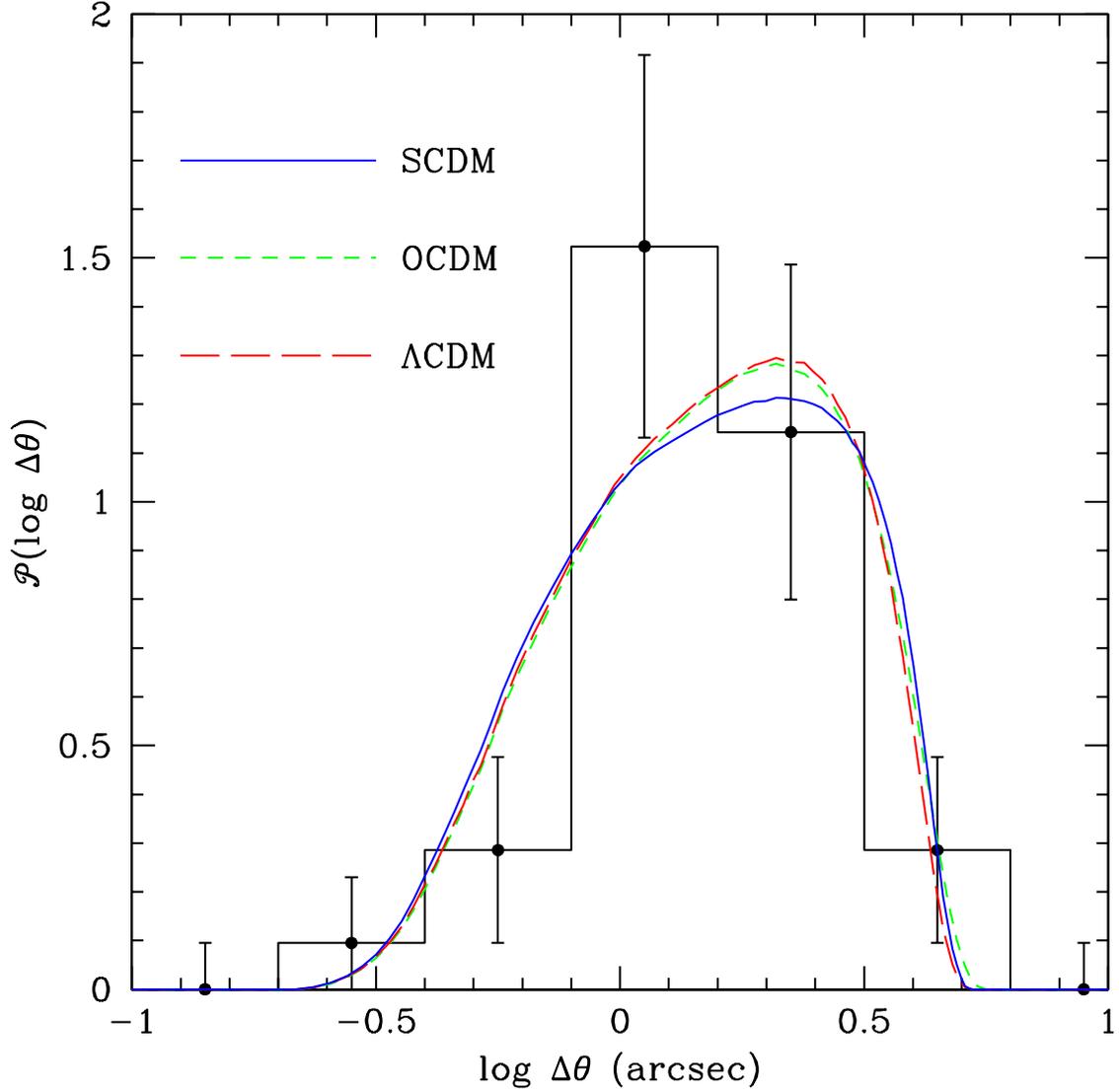}
\caption{Predicted probability distribution of angular separations between the 
two brightest
images of a lensed point source at $z_s=2$ in different cosmologies. 
The lenses 
are approximated as singular isothermal spheres for $M<3.5\times 10^{13}
\,M_\odot$, as Navarro-Frenk-White halos otherwise, and are distributed in 
mass according to the Press-Schechter theory. The angular selection effects
have been included following Kochanek (1999, private communication). 
The data with errors bars are taken from the CASTLE survey.    
}
\label{fig0}
\end{figure}

As suggested by Keeton (1998), a plausible way to reconcile simple statistical 
lensing models with the data on image separations of QSOs
is to consider the 
dissipation and cooling of the baryonic protogalactic component  
and the radial re-distribution of the collisionless dark matter as 
a consequence of baryonic infall (e.g. Blumenthal \etal 1986).
The net effect is a compression which transforms galaxy sized NFW halos 
into nearly isothermal distributions.
Group and cluster halos instead maintain their NFW profile,
and are extremely inefficient lenses compared to the corresponding SIS.
Since baryonic cooling is effective in halos having masses
smaller than a threshold value $M_c$ roughly independent of $z$ (Rees \& 
Ostriker 1977; Silk 1977), the picture adopted in this paper comprises then 
a collection of isothermal spheres for $M<M_c$ and a set of NFW halos for 
$M\geq M_c$. We have selected the value of $M_c$ empirically, by using the 
Kolmogorov-Smirnov test to optimize the agreement with the data on angular 
separations available in the CASTLE (CfA-Arizona-Space-Telescope-LEns) survey.
As shown in Figure \ref{fig0}, a `transition' mass of $M_c\simeq 3.5 \times 
10^{13} M_\odot$ provides a best-fit value for all the relevant cosmologies. 
As multiply imaged systems with large angular separations
are always produced by massive NFW halos, a reduction (with respect to the 
PS+SIS case discussed in the previous section) of the 
expected fraction of large separation systems to levels compatible with 
observations is obtained.

\subsection {Lensing optical depth in CDM cosmologies}

\begin{figure}
\plotone{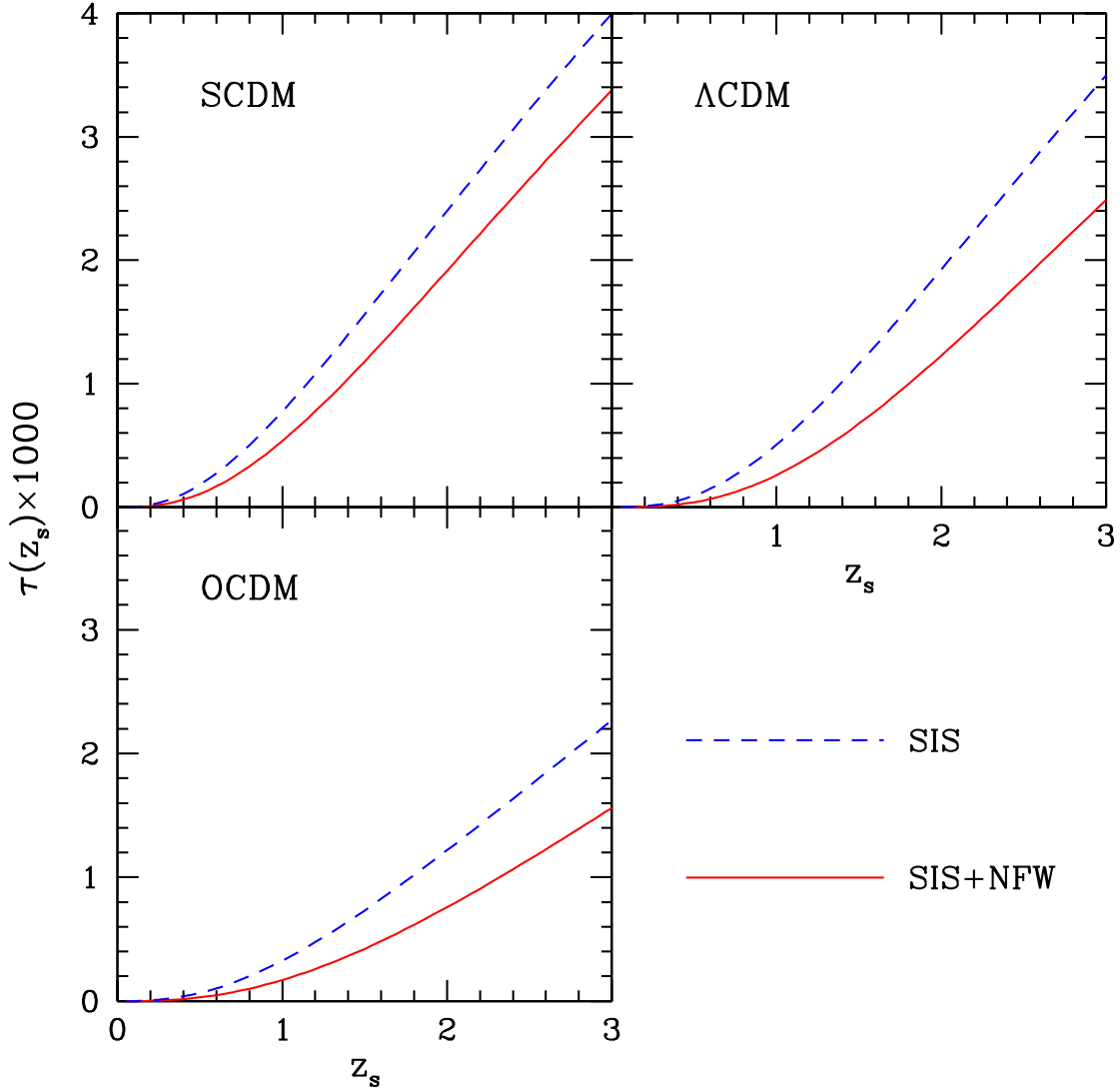}
\caption{Strong lensing optical depths for a point source at $z_s$ in three 
different hierarchical cosmogonies. The mass distribution of the lenses 
is described by the Press-Schechter theory.
{\it Solid lines:} a lens having mass $M$ is modeled by
a singular isothermal sphere for $M<3.5\times 10^{13} M_\odot$,
and by a NFW profile otherwise.
{\it Dashed lines:} each lens is modeled by a singular isothermal sphere.  
In this case the lensing optical depth for magnifications 
$\Delta m>0.3$ ($>0.1$) mag is 10 (100) times larger than plotted here. 
}
\label{fig1}
\end{figure}

\begin{figure}
\plotone{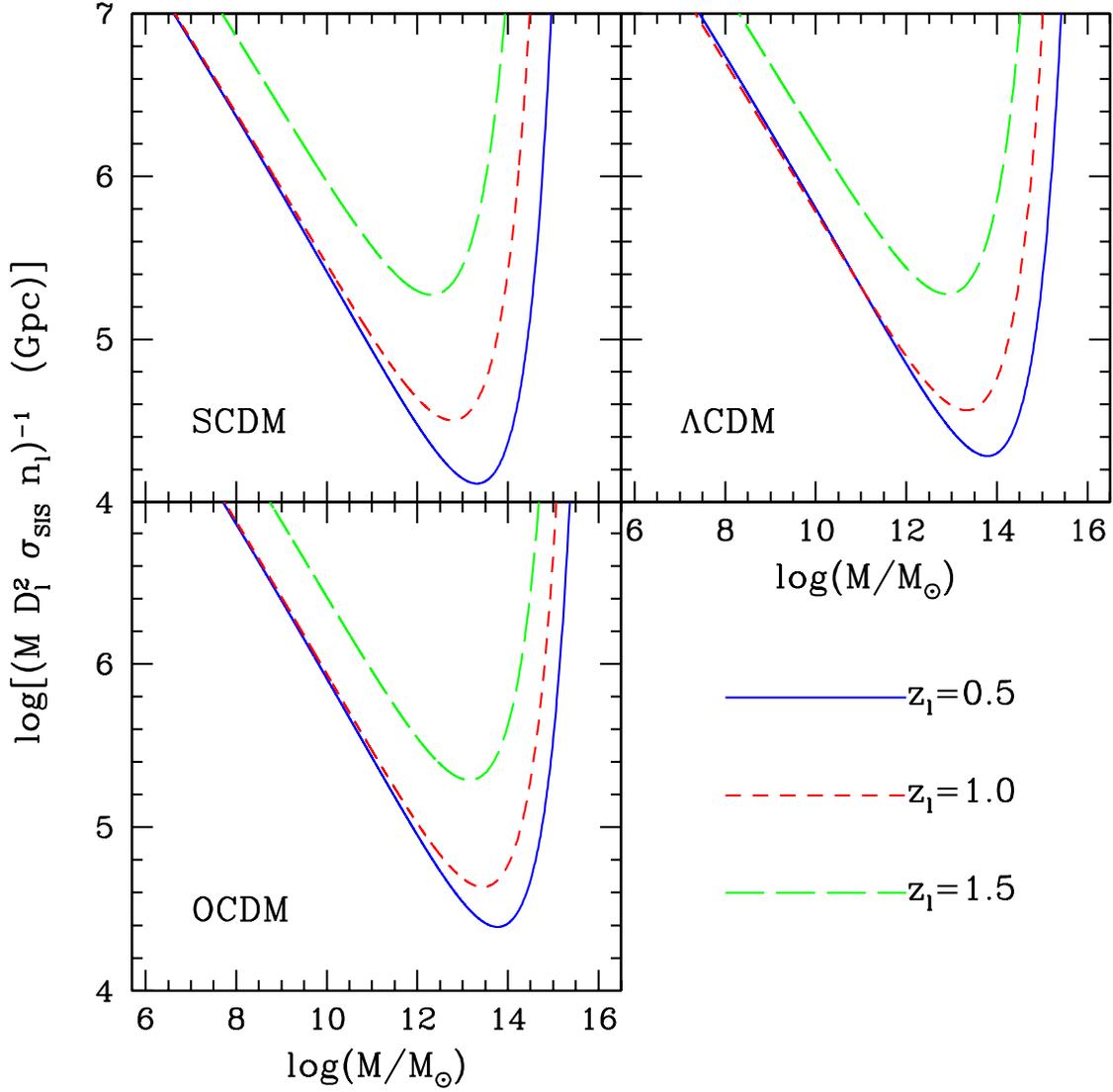}
\caption{Mean-free-path per logarithmic mass interval of a beam to 
encountering a strong lensing event, as a function of mass for different 
values of the lens 
redshift $z_l$ and for different cosmological models. The source is assumed to
be at $z_s=2$. Each lens is modeled by a singular isothermal 
sphere.}
\label{fig2}
\end{figure}

We analyze the frequency of lensing events in three different cosmologies with
parameters suggested by a variety of recent observations: an open model (OCDM, 
$\Om_M=0.3$, $\Om_\Lambda=0$), a flat model ($\Lambda$CDM, $\Om_M=0.35$, 
$\Om_\Lambda=0.65$), and  an Einstein-de Sitter model (SCDM, $\Om_M=1$, 
$\Om_\Lambda=0$).  In all cases the amplitude of the power spectrum or, 
equivalently, the value of the {\it rms} mass fluctuation in a $8 \; h^{-1}$ 
Mpc sphere, $\s_8$, has been fixed in order to reproduce the observed 
abundance of rich galaxy clusters in the local universe (e.g. Eke, Cole \& 
Frenk 1996; Jenkins \etal 1998).\footnote{Here and below we denote the 
present-day Hubble constant as $H_0=100 h\kmsmpc$.}~ 
The model parameters are shown in Table \ref{cosmo}.
The ``cosmic concordance'' model proposed by Ostriker \& Steinhardt (1995) has
been selected to represent the whole class of $\Lambda$-dominated cosmogonies 
($\Lambda$CDM). While this model is able to account for most known 
observational constraints, ranging from globular clusters ages through CMB 
anisotropies to recent analyses of the Type Ia SNe Hubble diagram, it may be 
in marginal conflict with 
the statistics of gravitational lenses (Kochanek 1996). Our choice of an open 
CDM (OCDM) cosmology is guided by the growing evidence in 
favour of a low value of $\Omega_M$ from cluster studies (White \& Fabian 1995; 
Carlberg, Yee, \& Ellingson 1997). For reference, and as instructive 
counterpart to the 
previous variants, we also consider a standard CDM (SCDM) scenario.
\begin{deluxetable}{l l l l l l}
\footnotesize
\tablenum{1}
\tablecaption{Cosmological parameters.  \label{cosmo}}
\tablewidth{0 pt}
\tablehead{
\colhead{Model} & $\Omega_M$ & $\Omega_\Lambda$ 
& $h$ & $n$ & $\s_8$ 
}
\startdata
SCDM  &  $1.00$ & $0.00$ & $ 0.50$ & $1.00 $ & $ 0.50 $\nl
OCDM  &  $0.30$ & $0.00$ & $ 0.70$ & $1.00 $ & $ 0.85 $ \nl
$\Lambda$CDM &$0.35$ & $0.65$ & $ 0.65$ & $0.96 $ & $0.87$ \nl 
\enddata
%\tablecomments{}
\end{deluxetable}
Figure \ref{fig1} shows the strong lensing optical depth
($\s=D_l^2 \s_{\rm SIS}$ for $M<M_c$ and $\s= D_l^2 \s_{\rm NFW}$ otherwise) 
for a source at redshift $z_s\le 3$ (also shown, for comparison, are the 
results obtained by assuming that every halo relaxes to a SIS). The lensing
frequency is the highest in SCDM 
and the lowest (by about a factor of 2 at $z_s=3$) in OCDM cosmologies, with 
$\Lambda$CDM models in between. 
In all cases $\tau$ is dominated by the contribution of
SIS lenses. The relative amplitudes between different cosmological
models depend on a 
combination of effects: ({\it a}) The geometry of the lens system affects 
the optical depth through the ratio $(D_l D_{ls}/D_s)^2 \equiv (c/H_0)^2 
g(z_l,z_s)$ and only halos in a narrow redshift interval can be efficient
lenses for a source at redshift $z_s$. If $z_s=2$ ($z_s=1$), for example, the 
maximum contribution to $\tau(z_s)$ comes from halos at $z\sim 0.5$ 
($z\sim 0.3$). For $z_l=0.5$ and $z_s=2$ 
(say), the ratios between the geometric factors in the various cosmologies 
are $g_{\rm SCDM}/g_{\Lambda{\rm CDM}}=0.58$ and $g_{\rm SCDM}/g_{\rm 
OCDM}=0.85$. ({\it b}) The path-length to 
a source at $z_s=2$  is (in units of the Hubble radius) $0.71$, $0.61$,
and $0.54$ in $\Lambda$CDM, OCDM, and SCDM universes, respectively.
 ({\it c}) For a fixed mass, the halo velocity dispersion shows only a 
weak dependence on the world model. At $z_l=0.5$, for example, 
$\sigma_v^4$ in $\Lambda$CDM is smaller than
in SCDM by $23\%$. On the other hand, OCDM shows the largest velocity
dispersions, with $\sigma_v^4$ larger than in SCDM by $6\%$. 
({\it d}) The number density of halos
obviously influences the lensing frequency. Roughly speaking, for objects with 
mass $M\ll M_\ast$, where $M_\ast$ is such that $\sigma_M(M_\ast)=\de_c(z)$, 
the number of halos contained in a Hubble volume is proportional to $\Omega_M$,
i.e. there are approximately three times more lenses in SCDM
than in low-density universes. By contrast, halos with $M\gta M_\ast$ form 
earlier in low-density cosmogonies. In SCDM, the formation of the massive (and 
exponentially rare) objects which are responsible for large separation 
multiple images is delayed to such low redshifts that they fail to be 
efficient lenses for sources at $z_s>1$. This is graphically shown in 
Figure \ref{fig2}, where the function $1/M D_l^2 \s_{\rm SIS}\, n_L(M,z_l)$,  
the mean-free-path per logarithmic mass interval of a beam to encountering a 
lensing event, is plotted as a 
function of mass for different values of $z_l$ and a source at $z_s=2$. While 
in the SIS approximation the   
lensing optical depth is dominated by masses in the range $10^{12}$ 
to $10^{14}\,\msun$, i.e. by massive galaxies, groups, and poor clusters,
a sharp drop in the mean free path is obviously present for $M>M_c$ 
in the combined SIS$+$NFW case.
The most efficient foreground lenses are at a redshift less than unity. 
The larger number of lenses in SCDM more than compensate for the longer 
path lengths to a given redshift and earlier collapse times that characterize 
open and $\Lambda$ models. 

It is fair at this stage to point out that, while it has become customary to 
model (as we have done) virialized dark matter halos by smooth distributions, 
observational data have recently shown the importance of considering the 
presence of baryonic and dark substructure in groups and clusters of galaxies 
(e.g. Williams, Navarro \& Bartelmann 1999; Keeton 1998). The relevance for 
lensing of substructure and, in particular, of massive central galaxies appears
to be more cospicuous in smaller aggregates.
This helps explaining why all the confirmed lens systems detected so far are
always produced either by galaxies (multiply imaged quasars and radio sources)
or by the halos of large clusters (arc systems).
Those lens systems which are clearly associated with a group or a small cluster
are always dominated by the lensing effect of a bright galaxy rather than
the dark halo of the group. While 
a detailed analysis of the effect of substructures is clearly beyond the scope
of this paper, we expect that our predictions for groups and small clusters 
may tend to underestimate their lensing optical depth (see also Williams 
\etal 1999).

It is also worth stressing that the PS mass function has been recently
shown to overpredict the number density of halos 
in the low-mass tail of the distribution by a factor
1.5--2 with respect to high-resolution $N$--body simulations
(e.g. Gross \etal 1998). 
In order to test how this inaccuracy of the PS formula affects
the computation of the strong lensing optical depth, we have re-calculated
$\tau(z_s)$ according to equation (\ref{tautog}) after replacing the PS mass 
function with the formula derived by Sheth \& Tormen (1999) in order to 
reproduce the $N$--body outcome. We find that, given the overall degree of 
uncertainty of our
calculations, the inaccuracy of the mass function does not represent
a major issue. For example, in the SCDM model and
for $z_s=2$, the strong lensing optical depth derived from $N$--body 
simulations is smaller than its PS counterpart by only 20\%. 

Another simplifying assumption we have made involves the spatial distribution 
of the lens population, here modeled with a Poisson process. Indeed, the 
clustering properties of dark halos are known to slightly alter the
lensing statistics. In particular, the presence of rich aggregates
and large voids will increase the frequency of events in the low- and 
high-magnification tails of the distribution (e.g. Jaroszy\'nski 1992).

\subsection{Galaxy lensing}

As a way to gauge the possible shortcomings of the PS method discussed above
[e.g. the intrinsic lack of accuracy at the low-mass tail 
as well as
uncertainties in $\de_c(z)$ or in the conversion between mass and 
velocity dispersion or concentration], 
it is interesting to consider in some detail the 
contribution of known galaxies to the magnification frequency. We model 
the lensing galaxies again as SIS. Their luminosity function, 
\be
n(L)= \f{n_\ast}{L_\ast} \left( \f{L}{L_\ast} \right)^\alpha
\exp \left(-\f{L}{L_\ast}\right)\;, \label{schect}
\ee
includes just early E/S0 types, as late spirals and irregulars are known 
to make a small contribution ($<20\%$) to the lensing statistics (Kochanek
1996; Maoz \& Rix 1993). The velocity dispersion of the dark matter in 
each galaxy halo is  related to the luminosity through a Faber-Jackson law 
\be
L=L_\ast \left( \f{\s_v}{\s_\ast}\right)^4\;.
\label{faja}
\ee
The  assumption that the local properties of E/S0 galaxies persist out to 
$z\sim 1$ has some observational basis (e.g. Lilly \etal 1995; van 
Dokkum \etal 1998). It also appears that, by and large, the statistics of 
multiply imaged QSOs is consistent with the locally observed galaxy 
population extrapolated to higher redshifts (Kochanek 1993; Maoz \& Rix 1993). 
If the galaxy comoving number density does not change with cosmic time,
and one neglects all evolutionary corrections to equation (\ref{faja}),
equation (\ref{tautog}) can be simplified, and the lensing 
optical depth becomes proportional to a dimensionless constant factor $F$ 
(Turner, Ostriker, \& Gott 1984) 
\be
F=\left({c\over H_0}\right)^3 \left({D_s\over 
D_{ls}}\right)^2\int_0^\infty dM \,
\s_{\rm SIS}(M,z,z_s) n_L(M,z)=16 \pi^3 n_\ast 
\left(\f{c}{H_0}\right)^3 \left( \f{\s_\ast}{c} \right)^4 \Gamma(2+\alpha)\;,
\label{fpar}
\ee
where $\Gamma(x)$ is Euler's gamma function. 
We use the best-fit E$+$S0 Schechter parameters from the SSRS2 samples  ($z\le
0.05$, Marzke \etal 1998): $\alpha=-1.00 \pm 0.09 $, $n_\ast=(4.4 \pm 0.8) 
\times 10^{-3} h^3\; {\rm Mpc}^{-3}$, and $M_\ast=-19.37\pm 0.1+5\log h\,$  
(the absolute 
blue magnitude associated with a luminosity $L_\ast$). With a fiducial value
of $\s_\ast=220\, \kms$ (Kochanek 1994), this model yields $F=0.017$ (cf. 
Fukugita \& Turner 1991).
For comparison, the all-type luminosity function from the Autofib survey 
($0.02<z<0.15$, Ellis \etal 1996), when scaled by the Postman \& Geller 
(1984) fraction (31\%) of E/S0 galaxies, gives $\alpha=-1.16 \pm 0.05 $, 
$n_\ast=(7.5\pm 1.3) \times 10^{-3} h^3\; {\rm Mpc}^{-3}$, $M_\ast=-19.30\pm 
0.13+5\log h$, and $F=0.033$. With these assumptions, cosmology only affects
the lensing optical depth through geometric effects, and the number of lensed
sources is the highest in a $\Lambda$CDM model. 
How does the optical depth
for strong galaxy lensing compare then with 
the PS+SIS+NFW frequency shown in Figure \ref{fig1}? 
For a source at (say) $z_s=2$, the former  (with Ellis \etal 
parameters) is 1.3 times larger than the latter in OCDM, 1.4 times larger  
in $\Lambda$CDM, and 2.9 times smaller in SCDM. The number of galaxy lenses 
is two times smaller with Marzke \etal parameters. 
In order to include the effects of galaxy groups and clusters, in the following
sections we shall estimate the number of magnified SNe from the PS 
formalism in a SCDM cosmology. It is clear from the above discussion, however, 
that the uncertainties in the lensing optical depth are non-negligible. 
One should also keep in mind that the PS mass function, $n(M)\propto 
M^{\zeta}$ with $\zeta\approx -2$, is significantly steeper at the low-mass end
than that derived from equations (\ref{schect}), and 
(\ref{faja}), $n(M)\propto M^{2\alpha+1}$. 
Barring selection effects, the statistics of gravitational lensing within 
the PS theory will tend to overestimate the frequency of small-separation 
images.   

\section{Supernovae}
\subsection{Rates}

The rates of SNe as a function of cosmic time depend on the star formation 
history of the universe, $\psi(t)$, the initial mass function (IMF) of 
stars, $\phi(m)$, 
and the nature of the binary companion in Type Ia events. The frequency of 
core-collapse supernovae, SN~II and possibly SN~Ib/c, which have 
short-lived ($\lta 50\,$ Myr) progenitors (e.g. Wheeler \& Swartz 1993), is 
essentially proportional to the instantaneous stellar birthrate 
of stars with mass $>8\msun$. For a Salpeter IMF (with lower and upper mass 
cutoffs of 0.1 and 125 $\msun$), the core-collapse supernova rate can be 
related to the star formation rate per unit comoving volume  according to 
\begin{equation}
R_{\rm II}(t)=\psi(t){\int_8^{125} dm \,\phi(m)\over \int_{0.1}^{125} 
dm \, m\, \phi(m)}=0.0074\times \left[{\psi(t)\over \sfrd} \right]\, \rd.
\end{equation} 

By contrast, the specific evolutionary history leading to a Type Ia event 
is poorly known. SN Ia are believed to result from the explosion of C-O
white dwarfs (WDs) triggered by the accretion of material from a
companion, possibly another WD or a non-degenerate, evolved star (e.g. 
Ruiz-Lapuente, Canal, \& Burkert 1997). The clock for the explosion is  
set by the lifetime of the primary star and, e.g., by the time taken for a 
non-degenerate companion to evolve and fill its Roche lobe (or, in the case 
the companion is another WD, the orbital decay time following gravitational
wave emission). The evolution of the rate may depend then, among other 
things, on the unknown mass distribution of the secondary binary components 
and on the distribution of initial orbital separations.  In 
this sense Type Ia SNe follow a slower evolutionary clock than Type II's, and 
their rate at any given time is sensitive to the past history of star 
formation in galaxies. Here we follow Madau, Della Valle, \& Panagia
(1998, hereafter MDP), and parameterize the rate of Type Ia's in terms of 
a characteristic explosion timescale, $\tau_{\rm WD}$ 
(which defines an explosion
probability per WD assumed to be independent of time), and an explosion
efficiency, $\eta$. The former accounts for the time elapsed in the various
scenarios from a newly born (primary) WD to the SN explosion itself, the 
latter for the fraction of stars in binary systems that will never undergo a 
SN Ia explosion because of unfavorable initial conditions. 
The rate of Type Ia events at any one time is then given by the explosions of 
all the binary WDs produced in the past that have not gone off already, i.e. 
\begin{equation}
R_{\rm Ia}(t)={\eta \int^t_0 dt' \,\psi(t') \int_{m_c}^8 dm \, 
\exp(-{t-t'-t_{\rm m}\over \tau_{\rm WD}})\phi(m) \over \tau_{\rm WD} \int dm \,m\,\phi(m)},
\end{equation}
where $m_c\equiv{\rm max}[3,m(t')]$, $m(t')=(10\,{\rm Gyr}/t')^{0.4}$ 
is the minimum mass of a star that 
reaches the WD phase at time $t'$, and $t_{\rm m}=10\,{\rm Gyr}/m^{2.5}$ is the 
standard lifetime of a star of mass $m$ (all stellar masses are expressed 
in solar units). 

\begin{figure}
\plotone{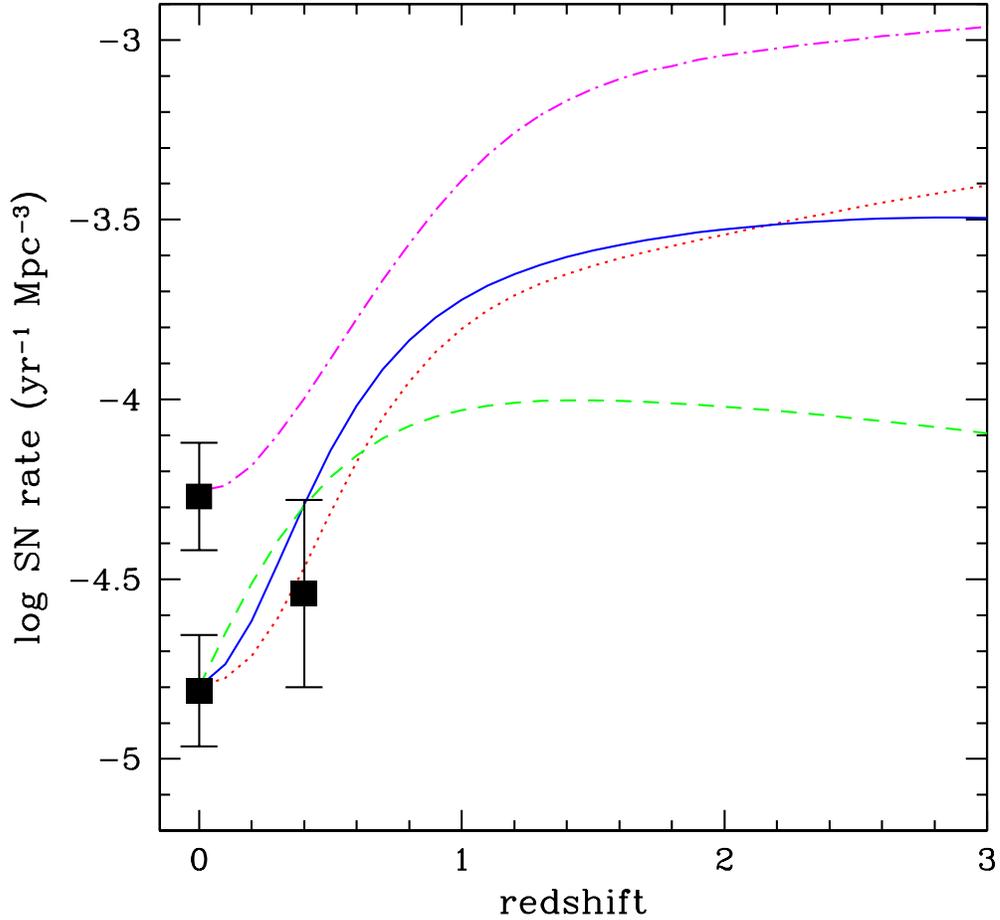}
\caption{Estimated Type Ia and II (rest-frame) frequencies as a function of
redshift. {\it Dot-dashed line:} SN II rate. 
{\it Dotted line:} SN Ia rate 
with $\tau_{\rm WD}=0.3$ Gyr. {\it Solid line:} SN Ia rate with $\tau_{\rm WD}=1$ Gyr. {\it 
Dashed line:} SN Ia rate with $\tau_{\rm WD}=3$ Gyr. The model assumes a 
SCDM cosmology. The data points with error
bars have been derived from the measurements of Cappellaro \etal (1997),
Tammann \etal (1994), Evans \etal (1989), and Pain \etal (1997). See 
MDP for details.
}
\label{fig3}
\end{figure}

Figure \ref{fig3} shows the predicted Type Ia and II rates per unit proper
time and unit comoving volume for a cosmic star formation history that traces
the rise from $z=0$ to $z\approx 1.5$ of the galaxy UV emissivity (Madau, 
Pozzetti, \& Dickinson 1998). The stellar evolution model (depicted in 
Figure 1{\it b}\ of MDP) assumes a Salpeter IMF and a weakly 
increasing star formation density above $z\approx2$, and is able to account 
for the entire optical background light recorded in the galaxy counts: half 
of the present-day stars -- the fraction contained in spheroidal systems -- 
formed at $z>2.5$. Because of the uncertainties in the amount of starlight that 
is absorbed by dust and re-radiated in the far-IR as a function of cosmic time,
these numbers are only meant to be indicative. Our estimates may be
on the conservative side, as the assumed stellar evolution model produces 
only a fraction $\sim 50\%$ of the IR background detected by COBE 
(Dwek \etal 1998). On the other hand, dust along the line-of-sight may affect
the detectability of Type II SNe (and also Type Ia's if the explosion delay
timescale is short enough that dust is retained in the star-forming regions.)
No attempt has been made to include the possibility of optically hidden SNe in
our calculations.

The Type Ia rates plotted in the figure assume characteristic ``delay'' 
timescales after the collapse of the primary star to a WD equal to 
$\tau_{\rm WD}=0.3, 1,$ and 3 Gyr, which virtually encompass most relevant possibilities.
For a fixed initial mass $m$, the frequency of Type Ia events peaks at an 
epoch that reflect an ``effective'' delay $\Delta t=\tau_{\rm WD}+t_{\rm m}$ from stellar 
birth. {\it A prompter (smaller $\tau_{\rm WD}$) explosion results in a higher SN Ia 
rate at early epochs.} The explosion efficiency ($5\lta \eta\lta 10\%$) was 
adjusted to reproduce the observed ratio of Type II to Type Ia rates in the 
local universe. When normalized to the emitted blue luminosity density, the 
predicted frequencies match rather well the data at $0\le z\le 0.4$ 
(MDP).   

\subsection{Number counts of SN Ia}

\begin{figure}
\plotone{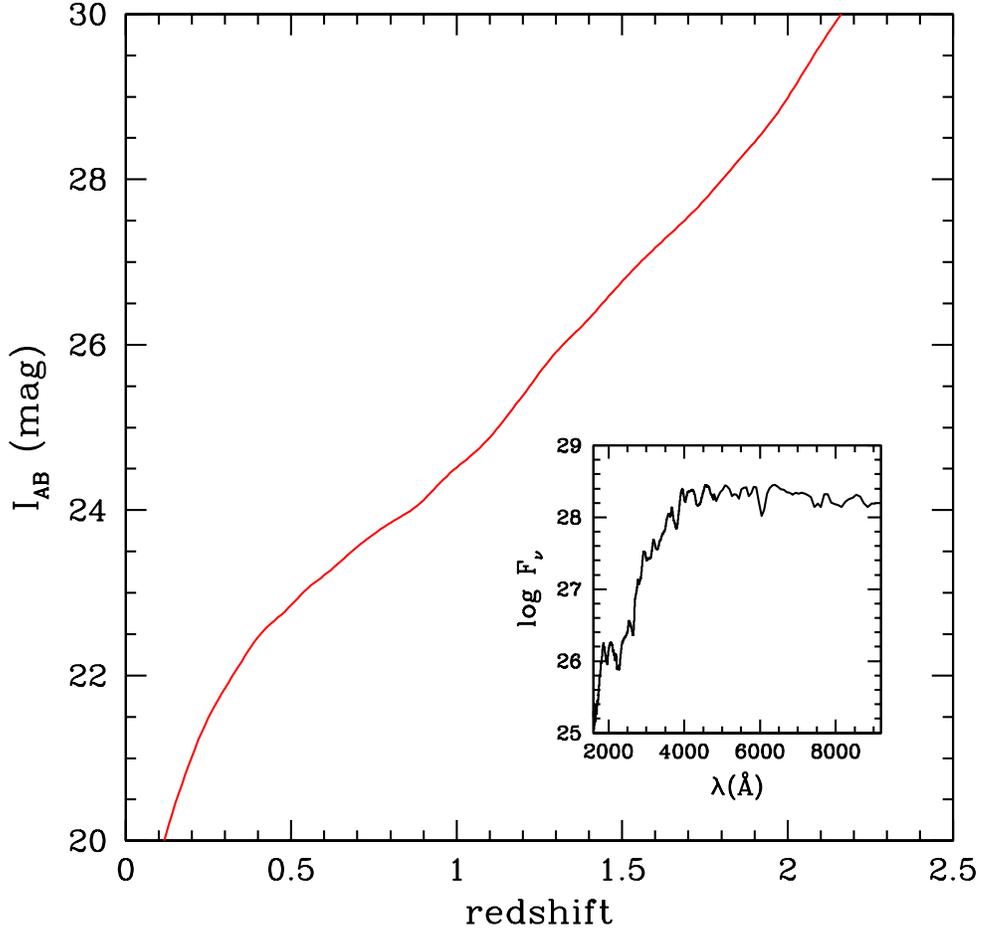}
\caption{$I$-band magnitude-redshift relation (SCDM) for Type Ia SNe. The 
apparent AB magnitude has been computed by redshifting and dimming the 
rest-frame $HST$--FOS spectrum of SN 1992A, and renormalizing it to 
$M_V=-19.4$. (The renormalized spectrum is shown in the inset, see Kirshner 
\etal 1993). The emitted power $L_\nu$ is in $\lumunits$.  
}
\label{fig4}
\end{figure}

In this section we will derive the observationally relevant quantity, 
i.e. the rate of significantly magnified SNe in a flux-limited survey, given 
a cosmological model, lens population, and star formation history. 
An apparent magnitude-redshift relation for unlensed SNe is needed
to estimate the number counts of distant Type Ia events. Traditional 
(i.e. from observed $B$- to emitted $B$-band light) $K$-corrections out to $z
\approx 0.2$ have been calculated by Hamuy \etal (1993) using available 
spectra of nearby SN Ia. Multi-band photometry has been used to extend these
results out to $z\approx 0.6$ (Kim, Goobar \& Perlmutter 1996). No accurate 
$K$-corrections have been determined, however, for supernovae at $z\ge 0.8$, 
as very few ultraviolet spectra from the local samples
with good signal-to-noise ratio are currently available.
To circumvent this problem we have constructed an (unlensed) magnitude-redshift
relation for Type Ia's by using as a template the $HST$--FOS spectrum of SN 
1992A taken 5 days after maximum blue light (Kirshner \etal 1993).
No correction for foreground and intrinsic extinction is apparently needed for 
this event (Burnstein \& Heiles 1984; Kirshner \etal 1993). The observed 
spectrum has been renormalized to a brightness (at maximum $B$-light) of
$M_V=-19.4$. The $m-z$ relation in the $I$-band is shown in Figure \ref{fig4} 
for $z<2.5$. Note that, 
because of the steep optical/UV spectrum of 1992A, our template Type Ia SN at 
$z=2.5$ will be much brighter in the near-IR ($J_{\rm AB}=26.9$) than in 
the optical ($I_{\rm AB}=31.2$). The $K$-correction significantly weakens the 
observed $I$ flux relative to the luminosity distance $D_{\rm L}^{-2}$ effect.  

For the intrinsic luminosity function of SN Ia, $\Phi_0(I,z)$, we have adopted 
a Gaussian distribution with standard deviation $0.15$ mag (cf. Van den Bergh 
\& McClure 1994). The luminosity function of lensed sources, $\Phi(I)$, can be 
obtained by convolving $\Phi_0(I,z)$ with the magnification probability
density (Vietri \& Ostriker 1983)
\be
\Phi(I,z)=\int dI' \,\Phi_0(I',z) \calP(\Delta m,z)  \;,
\label{prop}
\ee
where $\calP(\Delta m,z)$ is the probability of having magnification 
$\Delta m=I'- I$ for a source at redshfit $z$, normalized so that
$\int_{-\infty}^\infty \calP(\Delta m,z) d(\Delta m) = 1$.
The {\it observed} rate on the 
sky of Type Ia events is then 
\be
{dN\over dt}(<I_{\rm lim})=\int dz \,\f{dV(z)}{dz} \f{R_{\rm Ia}(z)}{1+z} 
\int_{-\infty}^{I_{\rm lim}} dI\, \Phi(I,z)\;,
\label{counts1}
\ee
where $V(z)$ is the comoving volume element surveyed, and the factor 
$(1+z)^{-1}$ accounts for the cosmological time dilation.\footnote{Note that, as
in a snapshot survey SNe are not necessarily detected at maximum light, the 
total number of events will actually depend on the effective
timescale over which a SN can be observed above threshold.}~From equation 
(\ref{prop}) we find 
\be 
{dN\over dt}(<I_{\rm lim})=\int dz \,\f{dV(z)}{dz} \f{R_{\rm Ia}(z)}{1+z} 
\int_{-\infty}^
{+\infty} dI \,\Phi_0(I,z) \int_{\mu_{\rm min}(I)}^{+\infty} d\mu \,
\calP(\mu,z) \;,
\label{counts2}
\ee
where $\log\, [\mu_{\rm min}(I)]=(I-I_{\rm lim})/2.5$ is the minimum 
magnification needed to detect a source in a flux-limited survey.
The above equation relates the number counts to the probability distribution 
of magnification discussed in \S~2 for SIS lenses. 
In the case of NFW halos, we first solve the lens equation for a set of values
of $k_s$ and of the dimensionless impact parameter $y=D_l \theta/r_s$. 
The magnitudes of the resulting images and the values of the corresponding 
impact parameters are stored in a file. The optical depth for a given 
magnification and source redshift is then computed using equation 
(\ref{tautog}): the suitable cross sections are calculated
by interpolating over the list of solutions of the lens equation,
and by using the method described in the appendix of NFW to evaluate 
the scale radius $r_s$ of a selected halo.

\begin{figure}
\plotone{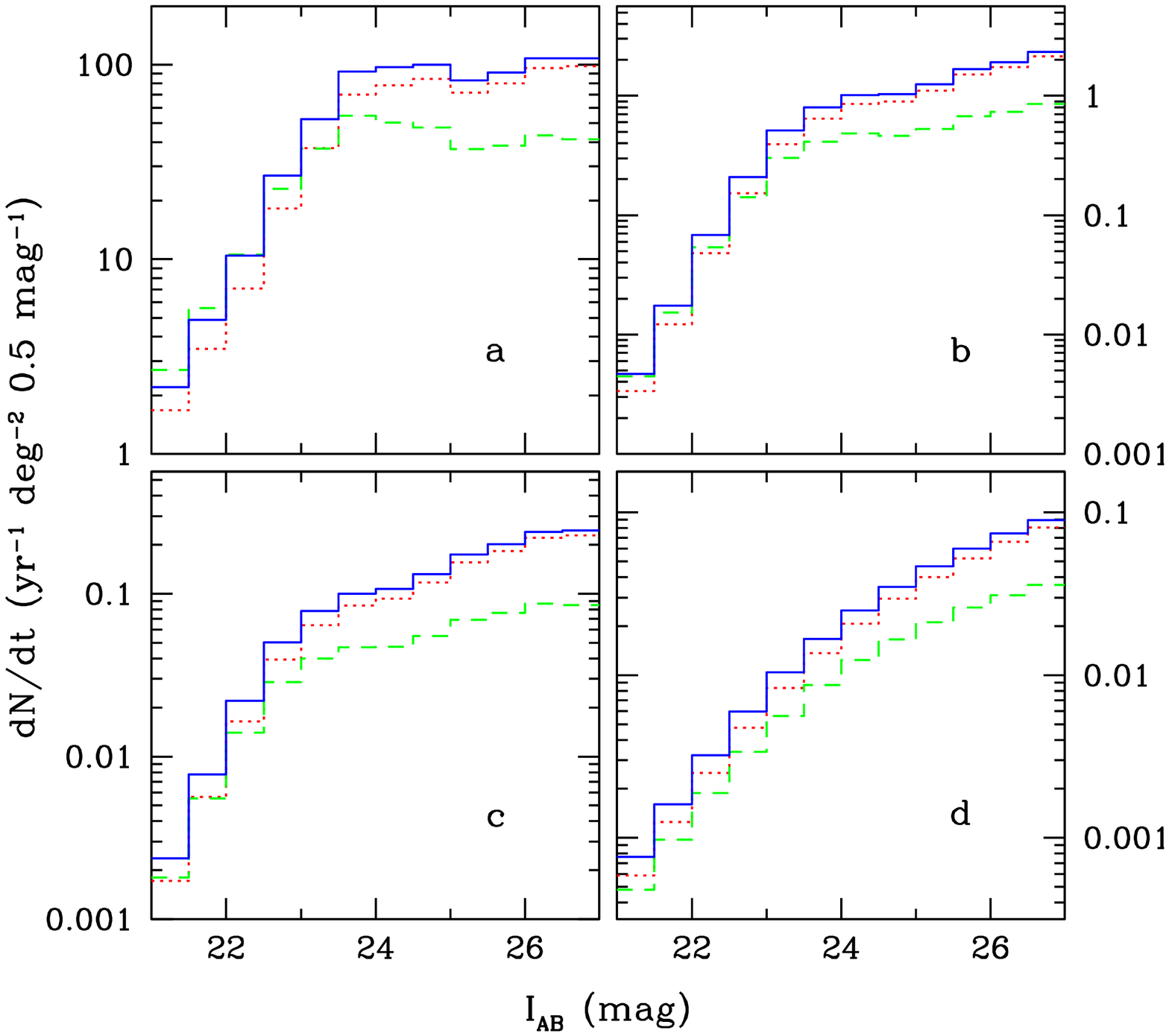}
\caption{Differential number counts of Type Ia SNe at maximum $B$-light
versus limiting $I$-band magnitude of the survey (SCDM universe).~
{\it a}) SNe in the absence of lensing; {\it b}) lensed SNe with 
magnification greater than 0.3 magnitudes (for strongly lensed sources only 
the brightest image is considered); {\it c}) 
lensed SNe with magnification greater than 0.75 mag (for strongly lensed
sources only the brightest image is considered); 
{\it d}) strongly lensed SNe with two images above the detection 
threshold (only the dimmest of the detectable images is considered).
{\it Dotted line:} rate assumes a time delay $\tau_{\rm WD}=0.3$ Gyr. 
{\it Solid line:} $\tau_{\rm WD}=1$ Gyr. {\it Dashed line:} $\tau_{\rm WD}=3$ Gyr. The 
effect of dust extinction on the detectability of Type Ia's has not been 
included in the models.
}
\label{fig5}
\end{figure}

To compute the total number of supernovae brighter than 
$I_{\rm lim}$ that have been gravitationally magnified by more than $\Delta m$ 
magnitudes, the function $\calP(\mu)$ in equation (\ref{counts2}) must be 
integrated from ${\rm max}\,(\mu_{\rm min}, 10^{\Delta m/2.5})$ to infinity.
Our treatment will be limited to magnifications $\Delta m\ge 0.3$ mag, i.e. 
to images that are more than 2$\sigma$ away from the mean intrinsic 
luminosity of SN Ia. In this regime, the weak (with typical magnifications of 
the order of a few per cent) lensing due to large scale structure can 
safely be neglected. 
Figure \ref{fig5} shows the counts of SNe Ia at maximum $B$-light for a SCDM 
cosmology.
The rate of detection in the absence of gravitational lensing effects 
(which is 
very close to the total expected number of events, since the probability 
of lensing is small) is compared with the estimated rates of multiply-imaged 
events (obtained by requiring that at least two images are sufficiently 
bright to enter the flux-limited sample; e.g., for SIS lenses, using eq. 
\ref{sigma-}), and with the rates of 
individual SNe magnified by more than 0.3 and 0.75
magnitudes (for strongly lensed SNe, only the brightest image is considered; 
i.e., for SIS lenses, $\calP(\mu)$ is computed 
using the cross section in 
eq. \ref{sigma+}, while, for NFW halos, the above mentioned 
numerical scheme is adopted).
Note that, for low magnifications ($\mu \lta 2$),
the lensing cross section of NFW halos is comparable (and generally larger) 
than that of the corresponding SIS halos.
Only considering multiply imaged systems, the relation $\sigma_{\rm NFW}\ll 
\sigma_{\rm SIS}$ holds.    
 
In a one square degree field, 
and for an effective observation duration of 1 yr, the total expected number 
of Type Ia SNe down to $I_\AB\le 25$ mag is (assuming $\tau_{\rm WD}=1$ Gyr) 
$\sim 390$. 
About 3.7 events will be gravitationally lensed by more than 0.3 
mag, 0.5 by more than 0.75 mag, and only 0.1 will have detectable double 
images.
These numbers are sensitive to the assumed explosion 
timescale.  Given a well-determined cosmic star formation history, it may 
be possible to infer something about the nature of Type Ia SNe from the 
observed number counts. It is also interesting to note that, in general,  
gravitational magnification facilitates the 
detection of SNe at higher redshift in a flux-limited sample, and
the fraction of lensed SNe tends to be higher at fainter 
magnitudes if the rate of events increases with look-back time and the gain 
in available volume is not balanced by a large $K$-correction. 
For $\Delta m \ge 0.75$, this is true 
up to $I_\AB\approx 23$. For fainter magnitudes, the increase in physical 
volume is smaller and the SN rate saturates (cf. Fig. \ref{fig3}); deeper 
searches will not lead then to an increased fraction of magnified sources. 
This defines an optimal 
survey sensitivity in the $I$-band for Type Ia's magnified by $\Delta m\ge 
0.75$ mag of $I_\AB\approx 23$ (cf. Linder \etal 1988). By contrast, there 
is no optimal flux threshold for detecting double images (Fig. \ref{fig5}d) 
in succession.

\subsection{Number counts of SNe II}

\begin{figure}
\plotone{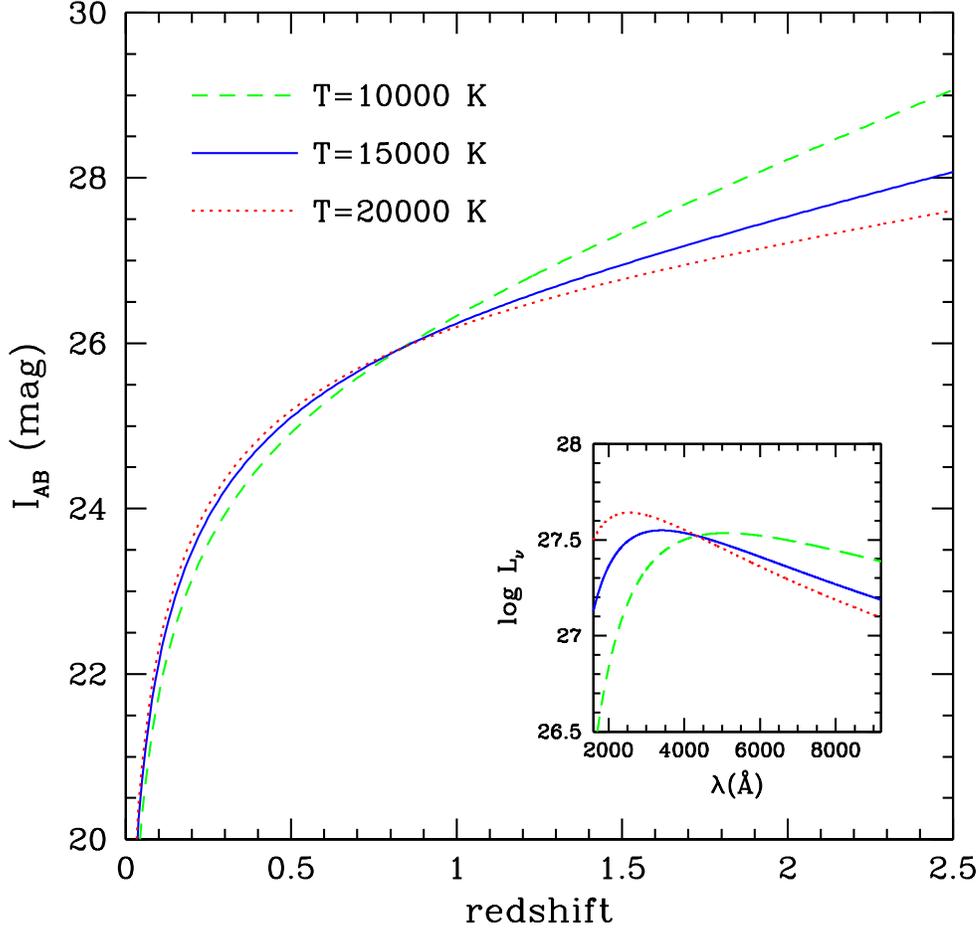}
\caption{Magnitude-redshift relation (SCDM) for Type II SNe. The apparent 
$I$ magnitude has been computed in the absence of dust reddening, and by 
assuming a blackbody spectrum at different temperatures. In all cases the 
emitted power has been renormalized to $M_B=-17.2$. {\it Dashed line}: 
$T=10000$ K. {\it Solid line}: $T=15000$ K. {\it Dotted line}: $T=20000$ K. 
}
\label{fig6}
\end{figure}

\begin{figure}
\plotone{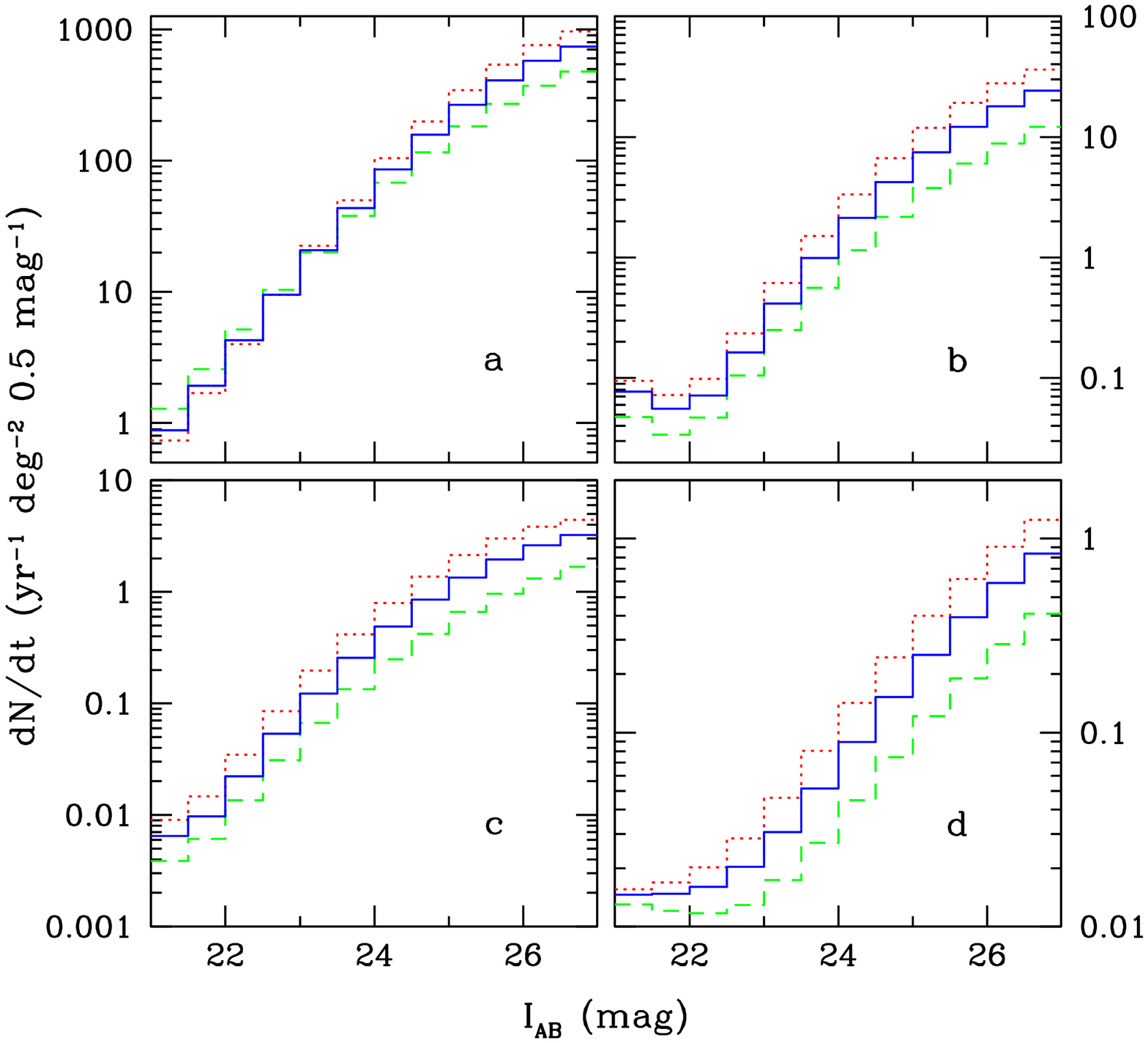}
\caption{Differential number counts of Type II SNe at maximum $B$-light
versus limiting $I$-band magnitude of the survey (SCDM universe, $z_{\rm max}=
5$).~
{\it a}) SNe in the absence of lensing; {\it b}) lensed SNe with magnification 
greater than 0.3 magnitudes (for strongly lensed sources only 
the brightest image is considered); {\it c}) 
lensed SNe with magnification greater than 0.75 mag (for strongly lensed
sources only the brightest image is considered); 
{\it d}) strongly lensed SNe with two images above the detection 
threshold (only the dimmest of the detectable images is considered).
{\it Dashed line:} rate assumes a photospheric temperature
of $T=10000$ K. {\it Solid line:} $T=15000$ K. {\it Dotted line:} 
$T=20000$ K. 
}
\label{fig7}
\end{figure}

Type II SNe are a couple of magnitudes fainter than SN Ia in the rest-frame
blue band.
Contrary to Type Ia's, they have a very wide luminosity function, which is
well represented by a broad Gaussian distribution
with $\langle M_B$(max)$\rangle=-17.2$ $(h=0.5)$ and
dispersion 1.2 mag (Tammann \& Schr\"{o}der 1990).\footnote{For the present
purposes, we shall ignore the possibility of an extended tail of Type II
events at bright (Patat \etal 1994) and faint (Woltjer 1997) magnitudes.}~ 
The spectra of most SN II at maximum light approximate well a 
single-temperature Planck function of about 15000 K from the UV through 
IR -- by contrast, Type Ia's exhibit a prominent UV deficit relative to
the blackbody fit at optical wavelengths (e.g. Filippenko 1997). The difference
in behavior of the $K$-correction is clearly visible in Figure \ref{fig6},
where the $I$-band $m-z$ relation for Type II's is plotted for three different
photospheric temperatures at peak. In the absence of dust reddening, the 
intrinsically fainter SN II appear brighter than SN Ia already at $z\approx 
1.5$. This has a strong effect on the expected number counts, shown in Figure 
\ref{fig7}. At $I_\AB\le 24$ mag, there is one SN II 
every 2.3 Type Ia's. The detection rate increases to one Type II every 1.2
SN Ia at $I_\AB \le 25$, and to one every 0.56 at $I_\AB \le 26$. 
Because of the flat 
$K$-correction and wide luminosity function, the fraction of lensed events is 
higher for Type II's, as only small magnifications are needed to bring 
high redshift events above the flux threshold (see also Linder \etal 1988).   

\subsection {Multiple images and time delays}

Although exceedingly rare in the field, multiply imaged Type Ia SNe are of 
particular interest for cosmography, as the propagation time from the source to 
the observer varies from one
image to another, and can be accurately measured if the source has a well
defined light curve. The time delay $\Delta t$ is proportional to the 
absolute scale of the lens geometry and hence inversely proportional to the 
Hubble constant (Refsdal 1964). Given a reliable time delay measurement,
a well-constrained lensing mass distribution (with a measured redshift and 
velocity dispersion), and a world model, one could estimate $H_0$.

\begin{figure}
\plotone{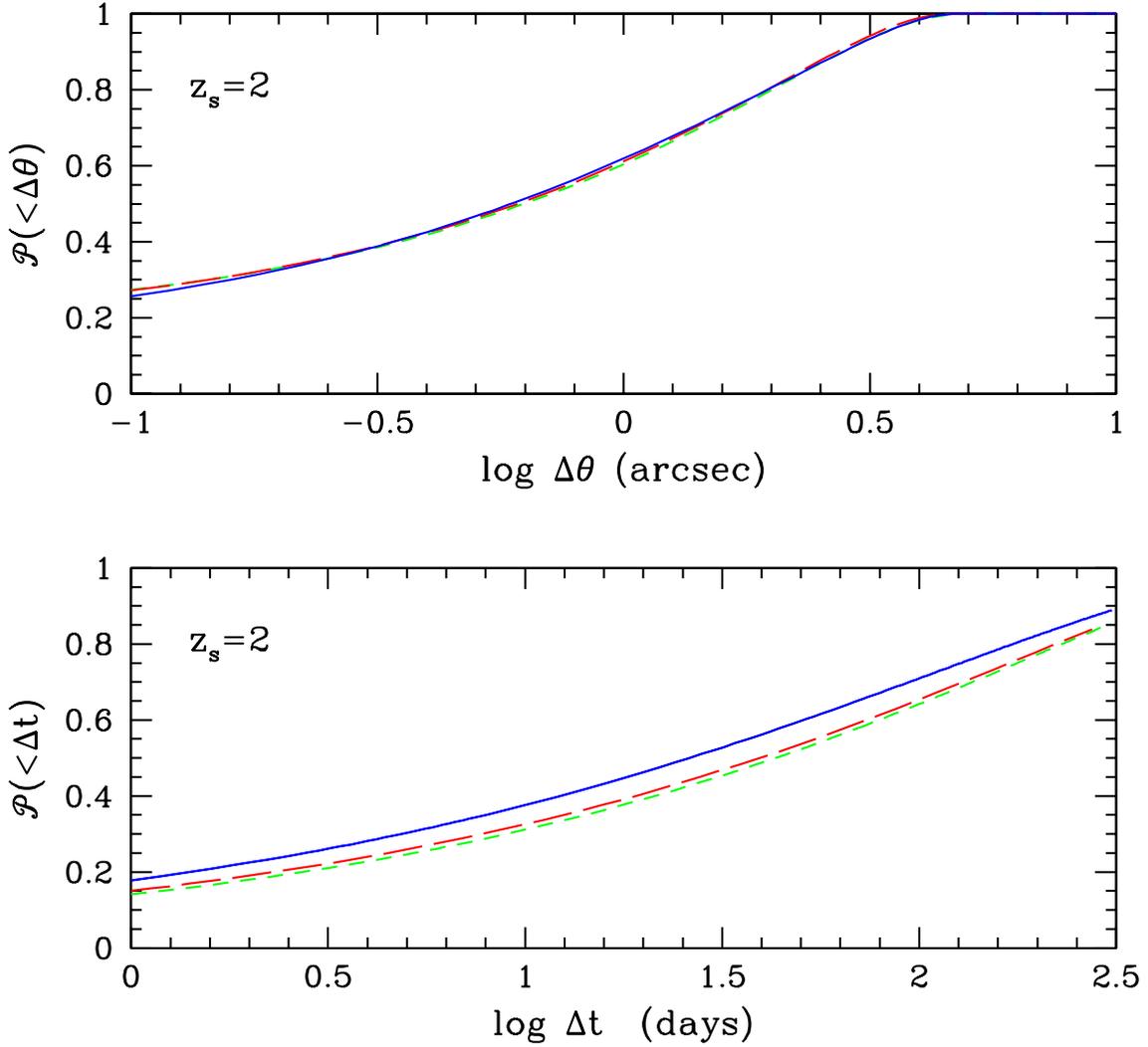}
\caption{Normalized cumulative probability distributions of angular separations
and time delays between the two brightest images of an SN at $z_s=2$. 
{\it Solid line:} 
SCDM. {\it Long-dashed line:} $\Lambda$CDM.
{\it Short-dashed line:} OCDM.} 
\label{fig8}
\end{figure}

The cumulative probability functions of image separations $\Delta\theta$ and 
time delays $\Delta t$ for a SN at $z_s=2$ in the field are shown in 
Figure (\ref{fig8}). About 60\% (SCDM) of double images will have angular 
separations $<1''$, and will therefore be flagged as multiple events in
ground-based surveys only if the time delay is comparable or larger than 
the SN decay time. The fraction of images with $\Delta\theta>4''$ is of 
order 1.5\%. Barring selection effects (see below), approximately
25\% of mirror events will be observed with an arrival time difference of 
$<3$ days, 40\% with $<10$ days, 50\% with $<30$ days.  
The frequency distributions have been computed  
by modifying equation (\ref{tautog}) to account for that fact that, at any 
redshift, only halos having mass larger than a threshold value can 
generate two images with a given angular separation or time delay. Normalizing 
the total probability to unity, and considering only SIS halos with $M<M_c$
(whose contribution dominates the strong lensing optical depth), one gets
\be
\calP(>\Delta\theta | z_s)=  \tau^{-1}(z_s) \int_0^{z_s} dz \,(1+z)^3
\f{dl}{dz}
\int_{{\cal M}(\Delta\theta,z,z_s)}^{M_c}
\!\!\!\!\!\!\!\! dM \,D_l^2 \s_{\rm SIS} \,n_L \;,
\label{theta}
\ee
where $\cal M$ is the mass that, for given $z$ and $z_s$, produces a 
separation $\Delta\theta$ [equal to $8\pi(\sigma_v({\cal M})/c)^2$ 
$(D_{ls}/D_s)$ for a SIS]. Similarly, one obtains
\be
\calP(>\Delta t | z_s)= \tau^{-1}(z_s) \int_0^{z_s} dz \, (1+z)^3
\f{dl}{dz} \int_{{\cal M}(\Delta t,z,z_s)}^{M_c} \!\!\!\!\!\!\!\!\!\! dM 
\left[1-\f{\Delta t^2}{\Delta t_{\rm max}^2(M,z,z_s)}\right]
\,D_l^2 \s_{\rm SIS} \,n_L \;,
\label{delt}
\ee
where $\Delta t_{\rm max}$ is the time delay at $\theta=\theta_E$ and 
$\cal M$ solves $\Delta t_{\rm max}({\cal M},z,z_s)=\Delta t$.
Note that, because of the small strong lensing cross section of NFW halos,
the presence of very large dark matter halos in OCDM and $\Lambda$CDM universes
does not create an extended tail towards large separations and long delays.

\subsection {Selection effects}

A peculiar feature of SNe Ia relative to QSOs is the relatively short lifetime 
of their emission.  While, for example, the finite angular resolution 
$\theta_{\rm res}$ of the observations may not allow to distinguish multiple 
images with separation on the sky $\Delta\theta<\theta_{\rm res}$,
if the time delay between different images is comparable or longer than the
typical decay time of a SN light curve [$t_{\rm SN}\sim 20 (1+z_s)$ days],
two mirror copies will appear roughly in the same position of the sky but at
different times. In this sense, even two images with $\Delta\theta<\theta_{\rm 
res}$ could in practice flag a strong lensing event if $\Delta t\gta  
t_{\rm SN}$: the same SN would go off in two (or more) incarnations of the 
same explosion, and the different images will be seen in succession. 
Because of the geometry of the source-lens-observer system, longer time delays 
will be obtained when the deflector is closer (in redshift)
to the radiation source.    
Consider, e.g, a SN at $z_s=2$ in a SCDM cosmology, lensed by a massive SIS 
halo ($M=1.2\times 10^{13}\, M_\odot$, $\sigma_v=250\,\kms$) at $z_l=1.2$.
Two images will result, with $\Delta\theta=0.83''<\theta_{\rm res}\sim 
1''$ (say). For a source-lens alignment $\theta=\theta_E/2$,
the magnifications of the images will be $\mu_+=3$ and $\mu_-=1$, and the
arrival time difference $\Delta t=81$ days. Unresolved images with shorter
delays may still be recognizable due to the large discrepancy between 
the apparent expected magnitude and the one observed, and a characteristic
{\it doubly-humped} light curve. At fixed $z_l$ and $z_s$, the 
angular separation 
between the lensed images depends only on the mass of the deflector, while the 
time delay is anti-correlated with the magnification: smaller delays correspond 
to more perfect source-lens alignments and thus to larger magnifications. 
Nearly synchronous images will then have rather similar brightnesses. 
A SN at $z_s=2$, lensed by a SIS halo at $z_l=0.5$ with mass 
$8.2\times 10^{12}\, M_\odot$ ($\sigma_v=220\,\kms$, SCDM), will produce 
when $\theta=0.15\,\theta_E$ two well resolved images 
($\Delta\theta=1.6''$), with magnifications $\mu_+=7.8$ and $\mu_-=5.8$. The
arrival time difference is only 20 days in this example.
For SIS lenses,
the maximum time delay occurs when $\theta=\theta_E$ ($\Delta t_{\rm 
max}\propto \theta_E^2$), hence less massive deflectors will always 
produce small-separation events and relatively short delays. 

\begin{figure}
\plotone{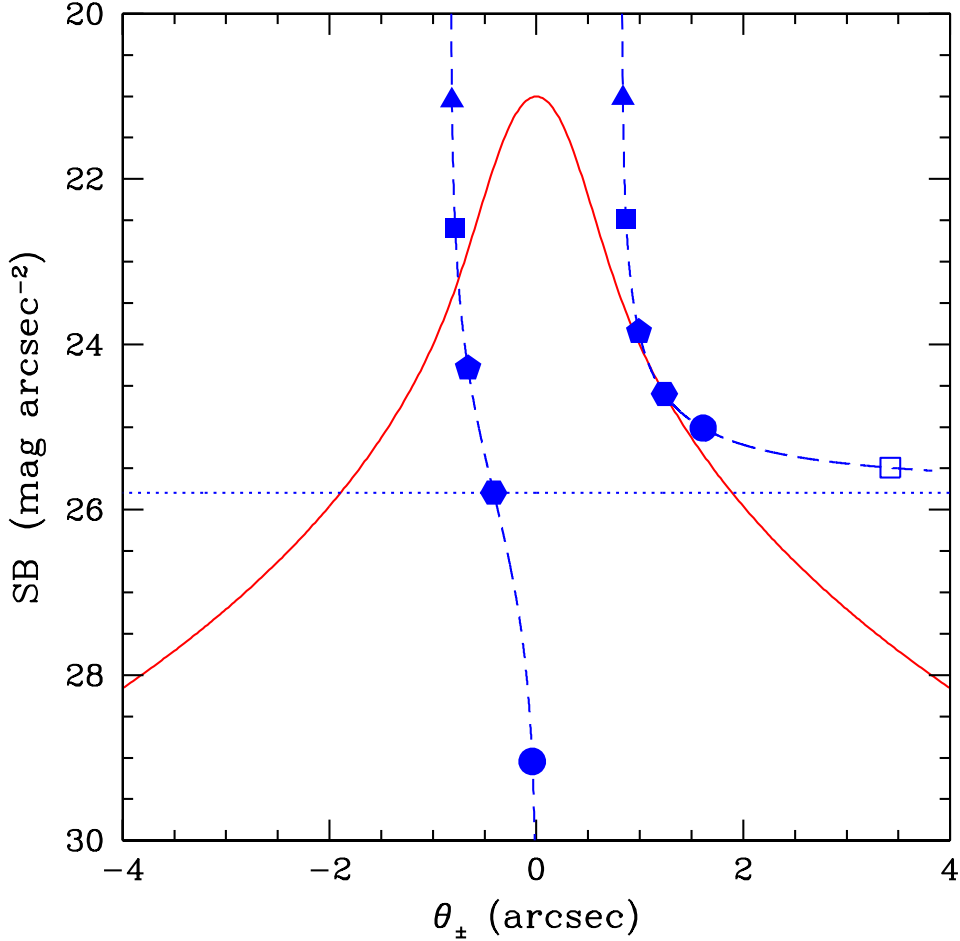}
\caption{Observed $I$-band light profile for an elliptical galaxy at 
$z_l=0.5$ (with typical luminosity and color, $M_B=-19.37+5\log h$, 
rest-frame $B-V=0.95$), as a function of angular distance from the galaxy 
center ({\it solid line}). A de Vaucouleurs' law has been adopted with 
effective radius $R_e=4$ kpc. The galaxy acts as a SIS lens with $\s_v=225\,
\kms$ for a background Type Ia SN at $z_s=1.5$ ($\theta_E=0.83''$). {\it Dashed 
lines:} peak surface brightness of the dimmest ($\theta_-=\theta-\theta_E$,
{\it left curve}) and brightest ($\theta_+=\theta+\theta_E$, {\it right 
curve}) SN images for different source-lens alignments $\theta$. The {\it 
filled points} along the curves identify multiple events corresponding to the
same impact parameter. Only one image is produced at $\theta_+>2\theta_E$. 
The {\it empty square} corresponds to a single image with $\Delta m=0.3$ mag.
The light profiles of SN and galaxy have been convolved with a 
Gaussian point spread function of FWHM $0.6''$. {\it Dotted line:}
SN brightness in the absence of lensing. A SCDM cosmology has been assumed. 
}
\label{fig9}
\end{figure}

A serious bias 
against the detection of small-separation images is due to the luminosity of 
the lensing galaxy. Galaxy-type halos with velocity dispersion 
$\s_v<200\,\kms$ 
completely dominate the rate of small-separation events. 
These must then be detected in the background of a lens elliptical galaxy, 
most probably at $z_l\sim 0.5$. Figure \ref{fig9} compares the observed 
surface brightness of a foreground $L_\ast$ 
elliptical with a background lensed Type Ia SN as a function of angular 
distance from the galaxy center. The effect of atmospheric seeing has been
approximated by convolving the true source flux with a Gaussian of dispersion
$\s={\rm FWHM}/\sqrt{8\ln 2}$. Even in the case of excellent FWHM$=0.6''$ 
seeing, the lens will greatly outshine the faintest SN image. Double events 
will only be detectable for nearly perfect lens-source alignments, an 
extremely rare case. In more modest seeing conditions (FWHM$=1''$), even 
the brightest 
image may remain undetected. From the ground, multiple SN events will be more 
easily observed in the background of a galaxy group or cluster. 

\begin{figure}
\plotone{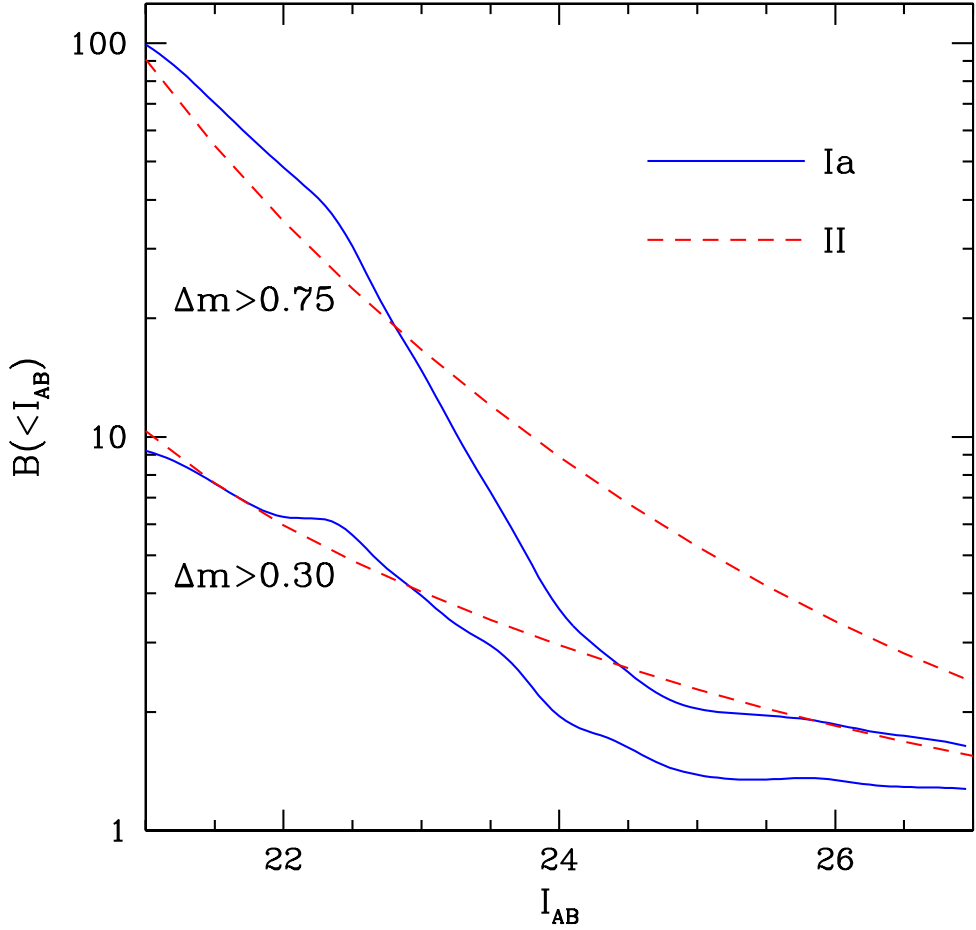}
\caption {Magnification bias versus apparent 
limiting magnitude.  {\it Solid lines:} Type Ia SNe 
with $\tau_{\rm WD}=1\,$Gyr.
{\it Dashed lines:} Type II SNe with  $T=15000\,$K. The two set of curves 
have been 
computed for magnifications $\Delta m \ge 0.3$ and $\Delta m \ge 0.75$ mag.
For strongly lensed sources, only the brightest image is considered.
}
\label{fig10}
\end{figure}

In a magnitude-limited sample, the sources that are observed to be 
gravitationally lensed include not only the lensed objects that are 
intrinsically brighter than the flux limit, but also sources that are 
intrinsically fainter and are brought into the sample by the magnification. 
Relative to the lensing optical depths shown in Figure \ref{fig1}, lensed 
systems will then be overrepresented among SNe of a given apparent magnitude. 
We have 
quantified this effect from the Type Ia and Type II number counts, by defining 
the magnification bias $B(<I_{\rm lim})$ as the ratio between the actual 
flux-limited counts of sources magnified by more than $\Delta m$ and the 
counts of lensed SNe that are intrinsically brighter than the flux limit,
\be
B(<I_{\rm lim})= \f{
\displaystyle{
\int dz \,\f{dV(z)}{dz} \f{R(z)}{1+z} 
\int_{-\infty}^
{+\infty} dI \,\Phi_0(I,z) \int_{\bar\mu}^{+\infty} d\mu \,
\calP(\mu,z)}
} 
{
\displaystyle{\int dz \,\f{dV(z)}{dz} \f{R(z)}{1+z} 
\int_{-\infty}^
{I_{\rm lim}} dI \,\Phi_0(I,z) \int_{10^{\Delta m/2.5}}^{+\infty} d\mu \,
\calP(\mu,z)}
}\;,
\label{bias} 
\ee
where $\bar \mu_{\rm }={\rm max}[\mu_{\rm min}(I), 10^{\Delta m/2.5}]$.
This definition generalizes that given in Turner, Ostriker, \& Gott (1984) 
and Fukugita \& Turner (1991) by taking into account the redshift dependence 
of the lensing optical depth.\footnote{In the notation of Fukugita \& Turner, 
our approach is equivalent to a flux-dependent lensing optical depth, 
as fainter objects are typically at higher redshifts than bright ones.}~
The magnification bias corresponding to $\Delta m=0.3$ and 0.75 mag is plotted 
in Figure \ref{fig10} for Type Ia ($\tau_{\rm WD}=1$ Gyr)
and Type II SNe ($T=15000$ K). At $I_\AB< 22$ the effect 
is huge, as the apparent frequency of events with $\Delta m \ge 0.75$ 
is boosted by a factor of $35-50$. 
Note that at these magnitude levels the counts are still nearly
Euclidean: the large bias is actually due to the rapidly increasing lensing 
frequency with lookback time. In the standard treatment with non-evolving 
optical depth, a large magnification bias is typically associated with integral 
number counts that increase faster than $f^{-2}$, where  $f$ is the observed 
flux (e.g. Schneider, Ehlers \& Falco 1992).

\subsection {Microlensing}

A modulation of the SN light curve may take place if, on top of the galaxy, 
group, or cluster magnification, a microlensing event occurs due to individual
stars or MACHOs in the galaxy halo or intracluster medium. The timescale for 
microlensing-induced variations is determined by two competing effects: the 
expansion velocity of the SN photosphere and the relative transverse
motion between the microlens and the light bundle. The radius of a SN at 
maximum light is $r_{\rm SN}\approx 10^{15}$ cm, corresponding  
to an observed angular dimension of $\theta_{\rm SN}=r_{\rm SN}/D_s$. A 
compact object along the line-of-sight to the SN will give rise to a detectable
time dependent amplification only when $\theta_E\gg\theta_{\rm SN} 
\gta |\theta|$ (Schneider \& Wagoner 1987), where $\theta$ is the angular 
impact parameter. Modeling the deflector with a Schwarzschild lens of mass 
$M$ (i.e. neglecting the shear induced by the gravitational field of 
the host halo), the first condition gives
\be
M\gg M_E=1.83 \times 10^{-4}\,M_\odot \;h \;\left(\f{c D_l}{H_0 D_s D_{ls}}
\right)\;.
\ee
Less massive MACHOs can efficiently magnify the SN only during its 
fast light-rising time. For $M\gg 
M_E$, the duration of the microlensing event roughly corresponds to the 
time needed by the photosphere to fill the Einstein ring of the lens 
(see also Gould 1994). For a typical photospheric expansion velocity of $\approx
10^9 \;{\rm cm \;s}^{-1}$, one obtains 
\be
t_{\rm ph}\approx 27\,{\rm days} \;h^{-1/2}\; \left( \f{H_0 D_s D_{ls}}{c D_l}
\right)^{1/2} \left( \f{M} {10^{-3} M_\odot} \right)^{1/2} (1+z_s)\;.
\ee
This is usually much shorter than the characteristic crossing time of the 
Einstein ring 
\be
t_E\approx 3.7\;{\rm yr}\;h^{-1/2} \left(\f{H_0 D_l D_{ls}}{c D_s}\right)^{1/2}
\left(\f{M}{10^{-3} M_\odot} \right)^{1/2}  \left(\f{200 \;{\rm km\, 
s}^{-1}}{v_\perp} \right) (1+z_l)\;, 
\ee
where $v_\perp \approx \sqrt{2} \s_v$ is the relative transverse velocity 
between microlens and source. In a SCDM cosmology with $z_s=2$ and $z_l=0.5$, 
one finds $t_{\rm ph}\approx 49.2 \;{\rm days}\, (M/10^{-3} M_\odot)^{1/2}$ and 
$t_E\approx 2.9\;{\rm yr}\, (M/10^{-3} M_\odot)^{1/2} (200 \;{\rm km \,s}^
{-1}/v)$. The microlensing timescale is then comparable with 
$t_{\rm SN}\approx 20 (1+z_s)$ days (SN Ia) for very small 
microlens masses. Only MACHOs with $10^{-4}\lta M/M_\odot \lta 10^{-3}$ can
then produce detectable `noise' that will degrade the light curves of the 
macroimages. More massive objects will boost the magnification of the SN 
image for the entire duration of the macroevent.

\section{Summary}

All previous studies of the influence that a population of deflectors has on
the observational properties of background objects have considered sources like 
galaxies or quasars that are quite inhomogeneous, i.e. have a wide range of 
intrinsic luminosities, ellipticities, etc. By contrast, Type~Ia SNe represent 
a population of objects with well-known intrinsic properties, and could -- 
in addition to their use as reliable cosmological distance indicators and 
valuable tracers of the star formation history in galaxies -- potentially  
flag the occurrence of a gravitational lensing event even in the absence 
of multiple or highly distorted images. We have investigated the lensing effect 
of background SNe due to virialized dark matter halos in CDM cosmogonies, 
and computed lensing frequencies, rates of SN explosions, and distributions of 
arrival time differences and image separations. Within the assumption that
dark halos approximate singular isothermal spheres on galaxy scales and NFW
profiles on group/cluster scales, and are distributed in 
mass according to the Press-Schechter theory, about one every 
12 Type Ia SNe at $z\approx 1$ will be magnified by $\Delta m\ge 0.1$
(SCDM). 
Using recent estimates of the global history of star formation to compute 
the expected frequencies of SNe as a function of cosmic time, we have 
derived a 
detection rate of Type Ia's with magnification $\Delta m\ge 0.3$ mag of a few 
events yr$^{-1}$ deg$^{-2}$ at maximum $B$-light and $I_\AB\le 25$. 
Events with $\Delta m\ge 0.75$ mag 
are 7 times less frequent. Only about one fifth of them give 
rise to observable multiple images: the SN will appear as multiple
events with identical light curves, separated in time and differing only by 
the scaling of their amplitudes. Because of the flat $K$-correction and 
wide luminosity function, we find that Type II SNe will dominate the number 
counts at $I_\AB>25$ and have the largest fraction of lensed objects. At 
bright magnitudes, the effect of magnification bias on the apparent frequency 
of lensed SNe with $\Delta m \ge 0.75$ is huge, a factor of $35-50$ in samples 
with $I_\AB \le 22$. 
The enhancement of the lensing probability beyond the optical depth expectation
drops to a factor of 3.5 at 24 mag (26 mag) in the case of Type 
Ia SNe (SN II). The apparent magnitude of the lens galaxy introduces a serious 
selection effect, as it reduces the detectability of small-separation multiple 
events in ground-based searches. 

On the face of their rarity, it is not unconceivable that large 
consortia of new and existing SN search teams may conduct in the near future 
deep pencil beam surveys and be able to detect magnified Type Ia's up to 
$z\sim 2$ (Wang 1998). Current searches have been limited so far to $R\sim 
24$ mag. A second-epoch deep ($I_\AB\sim 27$ mag) observation of the 
{\it Hubble Deep Field} by Gilliland, Nugent, \& Phillips (1999) has recently 
revealed a $z\sim 1.3$ Type Ia SN. The combination of a new generation of 
large telescopes with 1 degree field of view and multi-fiber spectrometers, 
together with faster techniques such as the snapshot distance method (Riess 
\etal 1998), could soon yield more than $\sim 500$ new events per year. 
As suggested by Kolatt \& Bartelmann (1998), one could also look for 
lensed SNe in the 
background of massive clusters. A nearby rich cluster at $z_l=0.1$ with 
virial mass of $10^{15} M_\odot$ and NFW profile 
will magnify by more than $0.3$ mag every point source
lying at less than $93''$ from its center.
The expected rate of SNe Ia behind
such a cluster with a mismatch of $\ge 0.3$ mag is then 
$\sim 1.2$ events per year with $I_\AB\le 26$ (SCDM, $\tau_{\rm WD}=1$ Gyr). 
Along the same lines, Kovner \& Paczynski (1988) have proposed the possible 
identification of Type II SNe in the lensed giant arcs observed in the cores 
of clusters of galaxies. If a SN goes off within the caustic and close to the
cusp of the cluster 
potential, it will be detected as three (or more) spatially-separated events 
with time delays as short as a few weeks or even days. The blue color of the 
arcs suggests significant star formation  hence a high rate of SN explosions.
Incidentally, if the cluster potential were 
understood well enough that a reasonable estimate of the time delay was  
available, one could exploit the occurrence of delayed mirror events to study
such transient phenomena in much greater details that would be possible 
without prior warning.  

Looking further into the future, it is interesting to consider the 
detectability of lensed Type Ia and Type II SNe with the {\it Next Generation 
Space Telescope} (NGST). A typical SN II at $z=5$ would give rise to an 
observed flux of 8 nJy (SCDM) at 2.6 \micron. At this wavelength,
the imaging sensitivity of an 8m NGST is around 1.5 nJy ($10^4$ s exposure and
$10\sigma$ detection threshold), while the moderate resolution ($\lambda/\Delta
\lambda=1000$) spectroscopic limit is about 50 times higher ($10^5$ s exposure
per resolution element and $10\sigma$ detection threshold, 
Stockman \etal 1998).
Depending on the history of star formation at high redshifts, the NGST could 
detect $\sim 10$ Type II SNe (at peak brightness) per $4'\times 
4'$ field per year at $z>4$ (MDP; cf. Dahl{\'e}n \& Fransson 1999;
Miralda-Escud\'e \& Rees 1997). (In principle, Type~Ia's could 
also be imaged at extreme redshifts, $z>10$, but the long delay times from 
stellar birth characteristic of these events make a detection 
rather improbable.) We find that, even for sources at these early epochs, the 
optical depth for strong lensing is small, $\tau\approx 0.008$ for $z_s=7$
and SCDM. Marri \& Ferrara (1998) have recently explored in details the 
observational perspectives
for Pop III SNe detection with the NGST when gravitational magnification is
taken into account. They computed lensing frequencies and magnification maps 
using a ray-shooting numerical technique and assuming point-like lenses. Our 
more realistic treatment implies probabilities that are much lower than 
theirs, as isothermal spheres and NFW halos are very inefficient magnifiers 
compared to point-like deflectors of the same mass. 

\acknowledgments
We have benefited from useful discussions with B. Bassett, 
R. Blandford, C. Keeton, N. 
Panagia, and P. Schneider. We thank R. Kirshner for providing us with 
the {\it HST}--FOS spectrum of SN 1992A, and the referee, C. Kochanek, for 
many comments that vastly improved the manuscript. Support for this work was 
provided by NASA through ATP grant NAG5-4236.

\references

Babul, A., \& Lee, M.~H. 1991, \mnras, 250, 407

Bartelmann, M. 1996, A\&A, 313, 697

Blandford, R.~D., \& Narayan, R. 1992, ARA\&A, 30, 311

Blandford, R. D., Saust, A. B., Brainerd, T. G., \& Villumsen, J. V. 1991,
\mnras, 251, 600

Blumenthal, G.~R., Faber, S.~M., Flores, R., \& Primack, J.~R. 1986,
\apj, 301, 27

Burnstein, D., \& Heiles, C. 1984, \apjs, 54, 33

Cappellaro, E., Turatto, M., Tsvetkov, D.~Yu., Bartunov, O.~S., Pollas, 
C., Evans, R., \& Hamuy, M. 1997, A\&A, 322, 431

Carlberg, R. G., Yee, H. K. C., \& Ellingson, E. 1997, \apj, 478, 462

Dahl{\'e}n, T., \& Fransson, C. 1999, A\&A, in press (astro-ph/9905201) 

Dwek, E., \etal 1998, \apj, 508, 106

Dyer, C.~C., \& Roeder, R. C. 1973, \apj, 180, L31

Ehlers, J., \& Schneider P. 1986, A\&A, 168, 57 

Eke, V.~R., Cole, S., \& Frenk, C.~S. 1996, \mnras, 282, 263 

Ellis, G.~F.~R., Bassett, B.~A., \& Dunsby, P.~K. 1998, Class. 
Quant. Grav, 15, 2345

Ellis, R. S., Colless, M., Broadhurst, T., Heyl, J., \& Glazebrook, K. 1996, 
\mnras, 280, 235

Evans, R., van den Bergh, S., \& McClure, R.D. 1989, \apj, 345, 752

Falco, E. E., Gorenstein, M. V., \& Shapiro, I. I. 1991, \apj, 372, 364

Filippenko, A. V. 1997, ARA\&A, 35, 309 

Flores, R., \& Primack, J.~R. 1996, \apj, 497, L5  

Frieman, J.~A. 1996, Comments Astrophys., 18, 323 

Fukugita, M., \& Turner, E. L. 1991, \mnras, 253, 99

Fukugita, M., Futamase, T., Kasai, M., \& Turner, E.~L. 1992, \apj, 393, 3

%Fukushige, T., \& Makino, J. 1994, \apj, 436, L111

Garnavich, P. M., \etal 1998, \apj, 493, L53

Gilliland, R. L., Nugent, P. E., \& Phillips, M. M. 1999, ApJ, 521, 30

Gould, A. 1994, \apj, 421, L71

Grogin, N. A., \& Narayan, R. 1996, \apj, 473, 570

Gross, M. A. K., Somerville, R. S., Primack, J. R., Holtzman, J., \&
Klypin, A. 1998, MNRAS, 301, 81

%Grossman, S. A., \& Narayan, R. 1988, \apj, 324, L37

Hamuy, M., Phillips, M.~M., Suntzeff, N.~B., Schommer, R.~A., Maza, J., 
\& Aviles, R. 1996, \aj, 112, 2391

Hamuy, M., Phillips, M.~M., Wells, L.~A., \& Maza, J. 1993, \pasp, 105, 787

Holz, D.~E. 1998, \apj, 506, L1

%Holz, D.~E., \& Wald, R. 1998, submitted to Phys. Rev. D (astro-ph/9708036)

Jaroszy\'nski, M. 1992, \mnras, 255, 655

Jenkins, A., \etal 1998, \apj, 499, 20

Kaiser, N. 1992, \apj, 388, 272

Kantowski, R., Vaughan. T., \& Branch, D. 1995, \apj, 447, 35

%Kasai, M., Futamase, T., \& Takahara, F. 1990, Phys. Lett., A147, 97

Keeton, C. R. 1998, PhD thesis, Harvard University

Kim, A., Goobar, A., \& Perlmutter, S. 1996, PASP, 108, 190

Kirshner, R.~P., \etal 1993, \apj, 415, 589

Kochanek, C.~S. 1993, \apj, 419, 12

Kochanek, C.~S. 1994, \apj, 436, 56

Kochanek, C.~S. 1995a, \apj, 445, 559

Kochanek, C.~S. 1995b, \apj, 453, 545

Kochanek, C.~S. 1996, \apj, 466, 638

Kochanek, C. S., Blandford, R. D., Lawrence, C. R., \& Narayan, R. 1989,
\mnras, 238, 43

Kolatt, T.~S., \& Bartelmann M. 1998, \mnras, 296, 763

Kovner, I., \& Paczynski, B. 1988, \apj, 335, L9

Kravtsov, A.~V., Klypin A.~A., Bullock J.~S., \& Primack, J.~R. 1998, \apj, 
502, 48 

Kundic, T., \etal 1997, \apj, 482, 75

Lacey, C., \& Cole S. 1993, \mnras, 262, 627
 
-------- 1994, \mnras, 271, 676

Lahav, O., Lilje, P.~B., Primack, J.~R., \& Rees, M.~J. 1991, \mnras, 251, 128

Lilly, S.~J., Tresse, L., Hammer, F., Crampton, D., \& Le F{\'e}vre,
O. 1995, \apj, 455, 108

%Linder, E.~V. 1998, \apj, 497, 28

Linder, E. V., Schneider, P., \& Wagoner, R. V. 1988, \apj, 324, 786

Lynds, R., \& Petrosian, V. 1989, \apj, 336, 1

Madau, P., Della Valle, M., \& Panagia, N. 1998, \mnras, 297, L17 (MDP) 

Madau, P., Pozzetti, L., \& Dickinson, M.~E. 1998, \apj, 498, 106

Maoz, D., \& Rix, H.-W. 1993, \apj, 416, 425

Maoz, D., Rix, H.-W., Gal-Yam, A., \& Gould, A. 1997, ApJ, 486, 75

Marri, S., \& Ferrara, A. 1998, \apj, 509, 43

Marzke, R.~O., da Costa, L. N., Pellegrini, P. S., Willmer, C. N. A., 
\& Geller, M. J. 1998, \apj, 503, 617

Metcalf, R.~B. 1999, \mnras, 305, 746

Miralda-Escud\'e, J., \& Rees, M. J. 1997, \apj, 478, L57

Moore, B., Quinn, T., Governato, F., Stadel, J., \& Lake G. 1999, submitted
to MNRAS (astro-ph/9903164) 

Narayan, R., \& Bartelmann, M. 1997, in Formation of Structure in the Universe,
ed. A. Dekel \& J. P. Ostriker (Cambridge: Cambridge University Press), 123

Narayan, R., \& White, S.~D.~M. 1998, \mnras, 231, 97P

Navarro, J. F., Frenk, C. S., \& White, S. D. M. 1997, \apj, 490, 493 

Ostriker, J. P., \& Steinhardt, P. J. 1995, Nature, 377, 600

Pain, R., \etal 1997, ApJ, 473, 356

Patat, R., Barbon, R., Cappellaro, E., \&  Turatto, M. 1994, A\&A, 282, 731

Perlmutter, S., \etal 1998, Nature, 391, 51

Postman, M., \& Geller, M. J. 1984, \apj, 281, 95 

Press, W.~H., \& Schechter, P. 1974, \apj, 187, 425

Rees, M.~J., \& Ostriker, J.~P. 1977, \mnras, 179, 541 

Refsdal, S. 1964, \mnras, 128, 295

Riess, A. G., Nugent, P., Filippenko, A. V., Kirshner, R. P., \& Perlmutter,
S. 1998, \apj, 504, 935

Riess, A. G., Press, W. H., \& Kirshner, R. P. 1996, \apj, 473, 88

Ruiz-Lapuente, P., Canal, R., \& Burkert, A. 1997, in 
Thermonuclear Supernovae, ed. P. Ruiz-Lapuente, R. Canal, \& J. Isern 
(Dordrecht: Kluwer), p. 205

Schneider, P., Ehlers, J., \& Falco, E.~E. 1992, Gravitational Lenses
(Berlin: Springer-Verlag)

Schneider, P., \& Wagoner, R. V. 1987, \apj, 314, 154

Sheth, R.~K., \& Tormen, G. 1999, submitted to MNRAS (astro-ph/9901122) 

Silk, J. 1977, \apj, 211, 638

%Soucail, G., \& Mellier, Y. 1994, in Gravitational Lenses in the Universe, 
%ed. J. Surdej, D. Fraipont-Caro, E. Gosset, S. Refsdal, \& M. Remy (Liege:
%Univ. Liege), 205

Stockman, H. S., Stiavelli, M., Im, M., \& Mather, J. C. 1998, in  ASP Conf. 
Ser. 133, Science with the Next Generation Space Telescope, ed. E. Smith \& 
A. Koratkar (San Francisco: ASP), p. 24

Tammann, G.~A., L\"offler, W., \& Schr\"oder, A. 1994, ApJS, 92, 487

Tammann, G.~A., \& Schr\"oder, A. 1990, A\&A, 236, 149

Tomita, K. 1998, Prog. Theor. Phys., 100, 1

Tremaine, S., Richstone, D.~O., Byun, Y.-I., Dressler, A., Faber, S.~M.,
Grillmair, C., Kormendy, J., \& Lauer, T.~R. 1994, \aj, 107, 634

Turner, E.~L., Ostriker, J.P., \& Gott, J.~R., III 1984, \apj, 284, 1

Tyson, J. A., \& Fisher, P. 1995, \apj, 446, L55

Tyson, J. A.,  Kochanski, G. P., \& Dell' Antonio, I. P. 1998, \apj, 498, L107

Van den Bergh, S., \& McClure, R.~D. 1994, \apj, 425, 205

van Dokkum, P. G., Franx, M., Kelson, D. D., \& Illingworth, G. D. 1998,
\apj, 504, L17

Vietri, M., \& Ostriker, J.~P. 1983, \apj, 267, 488

Villumsen, J. V. 1996, \mnras, 281, 369

Wallington S., \& Narayan R. 1993, \apj, 403, 517

Wambsganss, J., Cen, R., Xu, G., \& Ostriker, J. P. 1997, \apj, 475, L81

Wang, Y. 1998, submitted to MNRAS (astro-ph/9806185)

%Watanabe, K., \& Tomita, K. 1990 \apj, 355, 1

Wheeler, J. C., \& Swartz, D. A. 1993, \ssr, 66, 425

White, D. A., \& Fabian, A. C. 1995, \mnras, 273, 72

Williams, L.~L.~.R., Navarro, J.~F., \& Bartelmann, M., submitted to ApJ 
(astro-ph/9905134) 

Woltjer, L. 1997, A\&A, 328, L29

Zel'dovich, Ya.~B. 1964, Soviet Astron. AJ, 8, 13

\vfill\eject

\end{document}